\documentclass[12pt]{iopart}
\usepackage{graphicx}

	\newcommand{\vet}[1]{\mathbf #1 }

\begin{document}

\title[Emergent ultrafast phenomena in correlated oxides and heterostructures]{Emergent ultrafast phenomena in correlated oxides and heterostructures}

\author{M. Gandolfi$^{1,2,3}$, L. Celardo$^{1,2}$, F. Borgonovi$^{1,2}$, G. Ferrini$^{1,2}$, A. Avella$^{4,5,6}$, F. Banfi$^{1,2}$, C. Giannetti$^{1,2}$}

\address{$^1$Department of Physics, Universit\`a Cattolica del Sacro Cuore, Brescia I-25121, Italy}
\address{$^2$i-LAMP (Interdisciplinary Laboratories for Advanced Materials Physics),
Universit\`a Cattolica del Sacro Cuore, Brescia I-25121, Italy}
\address{$^3$Laboratory of Soft Matter and Biophysics, Department of Physics and Astronomy, KU Leuven, Celestijnenlaan 200D, B-3001 Heverlee, Leuven, Belgium}
\address{$^4$Dipartimento di Fisica "E.R. Caianiello", Universit\`a degli Studi
di Salerno, I-84084 Fisciano (SA), Italy}
\address{$^5$CNR-SPIN, UoS di Salerno, Via Giovanni Paolo II 132, I-84084 Fisciano
(SA), Italy}
\address{$^6$Unit\`a CNISM di Salerno, Universit\`a degli Studi di Salerno, I-84084
Fisciano (SA), Italy}

\ead{claudio.giannetti@unicatt.it, francesco.banfi@unicatt.it, avella@physics.unisa.it}
\vspace{10pt}

\begin{abstract}
The possibility of investigating the dynamics of solids on timescales faster than the thermalization of the internal degrees of freedom has disclosed novel non-equilibrium phenomena that have no counterpart at equilibrium. Transition metal oxides (TMOs) provide an interesting playground in which the correlations among the charges in the metal $d$-orbitals give rise to a wealth of intriguing electronic and thermodynamic properties involving the spin, charge, lattice and orbital orders. Furthermore, the physical properties of TMOs can be engineered at the atomic level, thus providing the platform to investigate the transport phenomena on timescales of the order of the intrinsic decoherence time of the charge excitations.  

Here, we review and discuss three paradigmatic examples of transient emerging properties that are expected to open new fields of research: i) the creation of non-thermal magnetic states in spin-orbit Mott insulators; ii) the possible exploitation of quantum paths for the transport and collection of charge excitations in TMO-based few-monolayers devices; iii) the transient wave-like behavior of the temperature field in strongly anisotropic TMOs.  
\end{abstract}

%
%
%
%
%

\section{Introduction: emergent phenomena at ultrafast timescales}
The dramatic advances in ultrafast science are boosting the development of a multitude of complementary techniques to investigate condensed matter systems strongly driven out of equilibrium \cite{Giannetti2016}. Nowadays, different excitation schemes can be used to selectively stimulate specific degrees of freedom: THz/infrared radiation can couple with the superconducting condensates and can be exploited to manipulate charge and spin gaps and to resonantly excite specific eigenmodes of the lattice; near infrared/visible light can be exploited to create \textit{non-thermal} fermionic and bosonic distributions and to artificially manipulate the band populations; UV/X-ray pulses resonate with charge-transfer processes or other optical transitions that involve deep electronic levels. At the same time, the multiple probing schemes currently available are concurring to the development of a coherent picture of ultrafast phenomena in solids. The complete reconstruction of the charge, spin and lattice dynamics triggered by the sudden photo-induced quench of solid state systems is close to realization. Besides the well-established possibility of probing the dynamics of the dielectric function in a broad range of the electromagnetic spectrum (from fractions of THz to the ultraviolet), the development of table-top and free electron laser (FEL) based X-ray sources is opening new tantalizing opportunities, ranging from the direct mapping of  
the occupation of high binding-energy electronic states via time-resolved X-ray photoemission to the reconstruction of the spin and lattice dynamics via resonant X-ray diffraction.

\begin{figure}
\begin{center}
\includegraphics[width=0.8\textwidth]{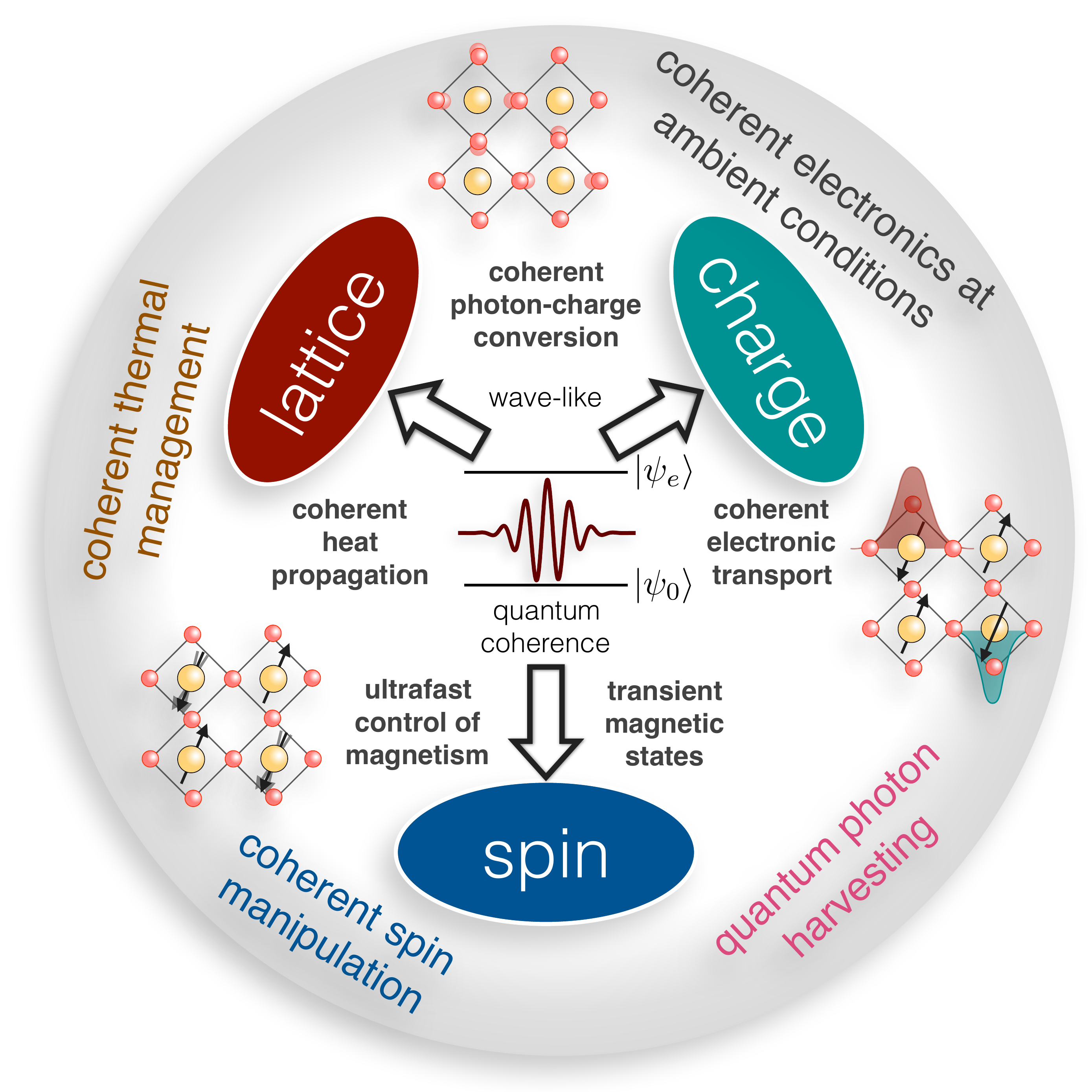} 
\caption{Cartoon representing the emergence of ultrafast properties and functionalities in correlated oxides}
\label{fig_intro}
\end{center}
\end{figure}
In principle, the possibility of creating \textit{artificial} initial states that evolve on a timescale shorter than the time necessary for the internal equilibration of the different fermionic and bosonic subsystems can be exploited to tackle many fundamental questions in condensed matter physics and to open completely unexplored fields. 
For example, cold atoms trapped into artificial optical lattices have been long exploited for investigating the way quantum many-body systems relax after an initial quench. The possibility of creating optical lattices without imperfections and time/spatial fluctuations (phonons) enables the study of \textit{ideal} solids in which the relevant interaction terms, such as the on-site Coulomb repulsion ($U$) and the  hopping ($t$) are selectively tuned. Furthermore, the small energy scales involved and the almost complete decoupling from the environment rise the typical relaxation timescales to milliseconds, thus allowing the real time investigation of the way many-body excited states propagate and eventually thermalize in an artificial interacting system that mimics a specific microscopic Hamiltonian. 

Whereas cold atoms provide a clean test bench for theories, they are hardly exploitable to explore the non-equilibrium physics of \textit{real} solids that constitute the building blocks for novel devices characterized by innovative functionalities. 
Among the many solid-state systems of technological interest, strongly correlated materials (SCM) $-$ in which the strong short-ranged electronic interactions provoke the failure of the independent-electron approximation $-$ provide a very interesting case-study for non-equilibrium physics. SCM are inherently characterized by the intertwining of the charge, spin and lattice degrees of freedom, which gives rise to an impressive number of physical processes that can be simultaneously accessed.
Nonetheless, the common perception is that the introduction of new variables $-$ the excitation pathway and the time evolution $-$ in the physics of correlated materials makes the treatment of these systems even more complex and pushes their modeling beyond any reliable theoretical scheme. In this review we will show that, in many interesting cases, the ultrafast capability of state-of-the-art techniques leads to the simplification of the physics of correlated materials and paves the way to the study of emergent phenomena that have no counterpart at equilibrium. This is particularly relevant at the nanometric scale and in few-atomic layers heterostructures, in which the timescales for the charge and energy propagation can be shorter than (or of the order of) the coupling with the thermal bath constituted by the spin and lattice fluctuations. In particular, transition-metal oxides (TMOs) are the ideal system for exploring these phenomena. As a consequence of the local correlations of the metal $d$ electrons, TMOs are always on the verge of multiple phase transitions. Furthermore, the possibility of controlling at the atomic level the construction of artificial blocks and heterostructures constitutes an additional knob that is expected to open exciting research fields.

Here, we will review and discuss three different paradigmatic examples of emergent phenomena at the ultrafast timescale:
\begin{itemize} 
\item[i] \textit{The magnetic dynamics in relativistic Mott insulators}.\\ 
$5d$ TMOs provide an interesting example of Mott insulators in which
the local moments have also orbital character. The dynamical constraints
related to the substantial spin-orbit coupling, to the peculiar characteristics
of $d$ electrons and to the actual topology of the lattice allow
to create transient non-thermal states and to better understand the
magnetic properties at equilibrium by photoinjecting charge excitations
that propagate in an artificial magnetic environment with no
counterpart at equilibrium.

\item[ii] \textit{Quantum-coherent phenomena in correlated oxide-heterostructures}.\\ 
When the size of the material is reduced to few atomic layers, the time necessary for collecting charge excitations is confined to the sub-10 fs timescale. This time is of the order of the decoherence time of the initial quantum wavefunction that is created by the absorption of a photon. We will argue that quantum-coherent effects, stemming from the propagation of local many-body excitations without loss of quantum-coherence, can be observed and potentially exploited on the ultrafast timescale in few-atomic-layers oxide heterostructures.    

\item[iii] \textit{Wave-like regime for the temperature propagation at the nanoscale}.\\
Describing the way the excess energy provided by an ultrashort light pulse propagates at the nanoscale is a very intriguing issue that challenges many well-established concepts. We will show that there exists a regime in which the temperature propagation exhibits an emergent wave-like behaviour that can impact the thermal transport properties of TMO nanostructures.   
\end{itemize}

\section{The magnetic dynamics in relativistic Mott insulators\label{iridium_oxides}}
\subsection{Optical control of the magnetization}
The capability
of inducing, detecting, and reversibly switching the magnetization of a material
on ultrafast timescales (from pico- to femto- seconds) and atomic distances is an extremely urgent task for material scientists \cite{Kirilyuk2010}. The
inescapable need and the ever-growing urgency are simply dictated by the daily-increasing technological demand to overcome the limitations,
in terms of speed and density, of today's magnetic memories and, more in general, digital electronics (logic gates, storage devices, ...).
This quest goes well beyond searching for a new material, which can optimize the actual technology, and regards the concurrent development of new ideas, tools and overall technologies to control the magnetic response of a functional material. One very promising route exploits ultrafast light pulses to photo-stimulate the emergence of novel magnetic states on timescales much shorter than the internal thermalization \cite{Balzer2015}. In particular, exploiting the interplay between the electronic and spin degrees of freedom, it is possible to induce transitions towards transient non-thermal states, whose magnetic properties are different from those exhibited in the equilibrium state. 
As an example, ultrafast optical techniques have been proposed as a tool to directly manipulate the exchange coupling $J_{exc}$ in Mott insulators \cite{Mentink2014,Mentink2015} and have been used to induce femtosecond magnetic switching in manganites \cite{Li2013} by exploiting an initial quantum-coherent magnetic regime on timescales faster than the coupling with the incoherent environment.
Moreover, the intrinsic capability of being reversible makes this all-optical routes usually preferable with respect to ordinary chemical or surface doping.
Along this line of thinking, it is urgent to find new classes of strongly correlated materials that can be efficiently used as playgrounds for inducing transient non-thermal magnetic states by means of ultrafast optical techniques.

\subsubsection*{\textbf{Spin-orbit Mott insulators: a new platform for ultrafast magnetism}.}
Mott insulators are always characterized by a very strong antiferromagnetic coupling driven by the virtual hopping ($t_{\mathrm{h}}$) of the electrons on neighbouring sites, with an energy cost equal to the on-site Coulomb repulsion $U$. The exchange energy in Mott insulators, which can assume very large values (up to $\approx$150 meV), is given by $J_{exc}$=2$t_{\mathrm{h}}^2/U$ and regulates the dynamics of the localized spins on the correlated sites. Because $J_{exc}$ originates from the electrostatic Coulomb repulsion and the Pauli principle, rather than on magnetic dipole forces, it can be directly modified by the change of the population of the Hubbard bands induced by photostimulation with ultrashort light pulses tuned across the Mott gap \cite{Mentink2014,Mentink2015}. 
\textit{Spin-orbit} Mott insulators are emerging as a new interesting class of magnetic correlated materials. In these systems, the strong spin-orbit coupling gives rise to Mott insulating phases in which the local moments have also orbital character \cite{Kim2008,Comin2012}, thus paving the way to the simultaneous control of both the magnetic state and the orbital occupation on the ultrafast timescale. 

The family of $5d$ transition-metal oxides, such as iridium oxides, is considered as one of the most promising systems to investigate the ultrafast charge/orbital/magnetic dynamics in  spin-orbit Mott insulators. 
In $5d$ TMOs one of the fundamental assumptions shared by almost all microscopical and phenomenological theories for strongly correlated systems $-$ the presence of a single energy scale dominating over all others $-$ dramatically fails. The increased average radius of the $5d$ orbitals, as compared to that of the $3d$ TMOs, simultaneously leads to the reduction of the on-site Coulomb
repulsion (down to $U\sim$1 eV) and to the increase of the hopping integrals (up to $t_{\mathrm{h}}\sim$0.4 eV) and, consequently, of the bandwidth $W$ \cite{Mazin2012}. As a consequence of the closeness of the $U$ and $W$ energy scales, neither a local Mott-like picture nor a delocalized Wannier scenario are fully appropriate to capture even the most basic physics of these compounds. 
As a further element, the large atomic number of the $5d$ metal ions makes the spin-orbit coupling
(SOC) extremely relevant (up to $\lambda_{SOC}\sim$0.7 eV) for the electronic and magnetic properties of the system \cite{Kim2008,Jackeli2009,Shitade2009}.
The topology of the lattice,
combined with the actual valence of the transition-metal ion, determines the orbital and spin effective interactions and, consequently, the possible orderings and critical temperatures of the system. As an example, the interplay of the lattice topology and spin-orbit coupling can lead to bond-directional magnetic interactions (Kitaev-like terms in the effective-spin Hamiltonian) \cite{Chun2015} that drive a strong frustration of the magnetic ground state and the possible emergence of exotic spin-liquid phases at low temperature \cite{Jackeli2009,Balents2010,Chaloupka2010,Chaloupka2013}.

In the following sections, we will provide two examples of $5d$ TMOs (the sodium iridate, Na$_2$IrO$_3$, and the strontium iridate, Sr$_2$IrO$_4$) in which the interaction between the SOC and the topology of the lattice links the dynamics of local magnetic moments to the creation, propagation and recombination of photo-injected charge excitations. The out-of-equilibrium approach permits not only to better understand the leading interactions that determine the equilibrium properties, but also to produce transient non-thermal states with an emergent magnetic dynamics due to the coupling of the photo-injected electron-hole excitations to an artificial magnetic environment with no counterpart at equilibrium. These findings open
 new routes in the investigation and, possibly, in the optical control of novel excitations and of exotic phases (e.g. spin-liquid) in $5d$ TMOs.

\subsection{Honeycomb-lattice sodium iridate}
In Na$_{2}$IrO$_{3}$, the edge-sharing IrO$_{6}$ octahedra form skewed (with respect to the crystallographic axes)
planes of Ir ions arranged in a honeycomb lattice (see Fig. \ref{fig_iridates}). The octahedral crystal-field of the oxygen cage splits the Ir-$5d$ orbitals in the two manifolds:
$e_{g}$, much higher in energy and out of the scope of the present analysis, and $t_{2g}$, at lower energy and occupied by $5$ electrons \cite{Felner2002,Kobayashi2003,Singh2010}. The actual distortions lift the degeneracy of the $t_{2g}$ manifold and affect the hopping amplitudes within and between the Ir hexagons \cite{Mazin2012,Mazin2013,Foyevtsova2013}. 
The system exhibits an insulating gap of $\sim$340 meV \cite{Comin2012} and quite complex magnetic properties: strong magnetic correlations below $\Theta_{corr}\sim$100
K and a long-range ordered zig-zag phase below $T_{\mathrm{N}}\sim$15 K \cite{Singh2010,Ye2012,Choi2012}. The large frustration parameter $\Theta_{corr}/T_{\mathrm{N}}\sim7$ suggests the tendency to form a low-temperature spin-liquid phase out of which the zig-zag order emerges.
The complexity of the electronic and magnetic properties is usually treated starting from two opposite perspectives, that rely on different starting point to construct the basis which describes the electronic wavefunctions.

\subsubsection*{\textbf{The fully localized picture}.} The fully-localized picture \cite{Kim2008,Chaloupka2010,Chaloupka2013} starts from the assumption that the SOC splits the six $t_{2g}$ levels and returns four completely-filled $J_{eff}=3/2$ levels and two half-filled $J_{eff}=1/2$ levels. This new arrangement of the levels opens up the possibility to get a single-band Mott insulator once the effect of even a small on-site Coulomb repulsion $U$ is properly taken into account. This $J_{eff}$ scenario leads to a description of the low-energy physics in terms of fully-localized effective moments whose dynamics is determined by a Kitaev-Heisenberg Hamiltonian \cite{Chaloupka2010,Chaloupka2013}. The structure of the Kitaev-like term and, in particular, its strength with respect to the one of the ordinary Heisenberg term strongly depends on the actual value of the SOC ($\lambda_{SOC}$ ) and on the specific topology of the lattice and determines the most favorable arrangement of the effective moments.
This issue is still open as the estimates present in the literature do not yet agree and lead to magnetic orderings not always compatible to the experimentally observed one.  The 340 meV insulating gap and the strong magnetic frustration, along with the recent measurement of bond directional interactions, are fully compatible with the $J_{eff}$
scenario that, at the moment, is the most accredited picture to describe the low-energy magnetic dynamics. 

\subsubsection*{\textbf{The delocalized picture: quasi molecular orbitals}.} If one takes into account
from the beginning the large oxygen-mediated hopping term ($\sim$250 meV) between neighboring Ir atoms, which drives a very fast electronic delocalization on the Ir hexagons, it seems reasonable to study the effects of the large SOC and of the small $U$ not on the local Ir-$t_{2g}$ levels, but on quasi-molecular orbitals (QMOs) extending over the whole Ir hexagon. The QMOs are the exact basis states of a non-interacting, zero SOC, isolated hexagon of Ir ions and are built from linear combinations of Ir-$t_{2g}$ Wannier functions \cite{Mazin2012,Foyevtsova2013}.
In terms of QMOs, the effective residual hopping is reduced to smaller values and to shorter extents better justifying a subsequent perturbative approach in terms of $\lambda_{SOC}$ and $U$. As a matter of fact, the QMO scenario not only perfectly rationalize the outcome of full-fledge
(actual distortions and lattice structure) relativistic band-structure calculations \cite{Mazin2012,Foyevtsova2013}, but it also accounts for the great variety of structures observed via photoemission and optical spectroscopy up to 2 eV \cite{Li2015a}, thus demonstrating its validity and efficiency.
\begin{figure}
\begin{center}
\includegraphics[width=1\textwidth]{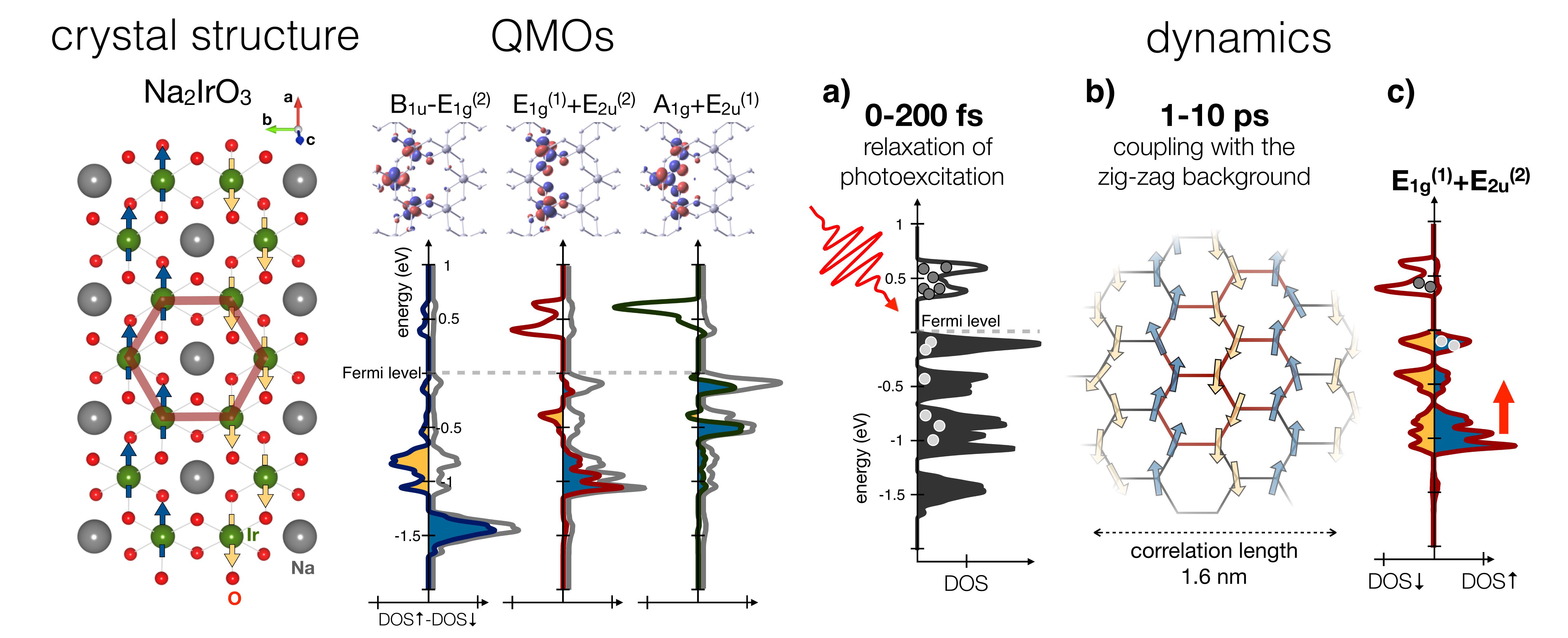} 
\caption{The central panel displays the honeycomb lattice structure of Na$_2$IrO$_3$. The magnetic zig-zag ordered phase is indicated by the blue (spin-up) and yellow (spin-down) arrows. The central panel shows the polarization of suitable combinations of QMOs, as calculated by ab-initio DFT calculations. The grey lines represent the total spin-down DOS. The polarization is defined as the differences between the QMO-projected spin-up and spin-down DOS. The three linear combinations shown in the figure reproduce both the spatial (see the wavefunctions reported in the top panel) and the magnetic polarization of the zig-zag ordered ground state. A cartoon of the relaxation processes
after the initial photoexcitation is reported in the right panels (a-c). The panel a) shows the
ab-initio total DOS of the occupied and unoccupied states.
The c) panel shows the spin-resolved DOS of the $E_{1g}^{(1)}\pm E_{2u}^{(2)}$ QMO combination.  Adapted from Ref. \cite{Nembrini2016}.}
\label{fig_iridates}
\end{center}
\end{figure}

\subsubsection*{\textbf{The non-equilibrium approach}.} In Ref.~\cite{Nembrini2016}, a non-equilibrium approach has been
adopted to investigate the interplay between the dynamics of the localized $J_{eff}$ moments and the delocalization of charge excitations on the Ir honeycombs (QMO picture).
 The possible link between the magentic dynamics and the QMO picture is directly provided by the Na$_{2}$IrO$_{3}$ electronic band structure (see Fig. \ref{fig_iridates}), that has been calculated by \textit{ab-initio} relativistic Density Functional Theory (DFT). As already observed in several works \cite{Mazin2012,Mazin2013,Foyevtsova2013,Li2015},
the DFT-calculated density of states (DOS) presents five separated bands which are strongly reminiscent of the QMOs. The six QMOs localized on a particular hexagon can be grouped into the lowest-energy $B_{1u}$ singlet, the two doublets $E_{1g}$ and $E_{2u}$ and the highest-energy
$A_{1g}$ singlet \cite{Mazin2012}. SOC mixes the three QMOs closer to the Fermi energy, splits the doublets and opens up a pseudogap \cite{Mazin2012,Foyevtsova2013} that widens up to the experimental insulating gap when the Coulomb repulsion $U_{eff}$ is considered \cite{Li2015a}. Interestingly, the QMO representation is fully compatible with the zig-zag ordered ground state observed at $T<T_{\mathrm{N}}$:
there exist three couples of QMOs that form linear combinations preserving both the spatial arrangement and the magnetic polarization of the zig-zag ordering. The $B_{1u}\pm E_{1g}^{(2)}$ modes are mostly confined to binding energies of the order of 0.5-1.5 eV, with a resulting very
small overlap with the states at the Fermi level, and feature a quite small overall net polarization. Instead, the $E_{1g}^{(1)}\pm E_{2u}^{(2)}$ modes, which are mainly located at $\sim$1 eV binding energy, are almost fully polarized according to the zig-zag pattern and are characterized
by a non-zero overlap with both the occupied states right below the Fermi level and the empty states at $\sim$+0.3 eV. In order to recover the full DOS of the low-energy occupied states, it is however necessary to consider also the contribution from the $A_{1g}\pm E_{2u}^{(1)}$
modes, which are fully polarized and extend in energy from $\sim$-0.3 eV to $\sim$+0.7 eV. The necessity of including more than one QMO combination to describe the valence $-$ lower-Hubbard band (LHB) in the localized description $-$ and the conduction $-$ upper-Hubbard band (UHB) in the localized description $-$ bands, suggests that the QMO picture, which captures the main features of the deep electronic states, is less efficient to describe the physics of the low-energy electronic excitations.

The $E_{1g}^{(1)}\pm E_{2u}^{(2)}$ QMO combination acts as perfect link between the low-energy dynamics and the high-energy electronic properties.
Considering its significant overlap with the low-energy states close to the Fermi level and the almost full polarization of both the occupied states at $\sim$-1 eV and the empty states at $\sim$+0.3 eV, it is natural to foresee a strong interplay between this QMO combination and electron-hole excitations photoinjected via ultrashort light pulses
and eventually decaying and releasing energy to the zig-zag magnetic background.
To test this scenario, Nembrini \emph{et al.} focused on the ultrafast dynamics of the optical conductivity in the 1.4-2.1 eV energy range, that corresponds to the optical transitions from the occupied $E_{1g}^{(1)}\pm E_{2u}^{(2)}$ and $B_{1u}\pm E_{1g}^{(2)}$ QMOs to the empty conduction band \cite{Nembrini2016}. The basic idea of the experiment is to use an ultrashort light pulse (photon energy 1.55 eV) to photoinject local electron-hole excitations across the Mott gap and to exploit the high binding energy QMOs to map the delocalization of the charges into the Ir honeycombs and the subsequent coupling of QMO-like charge excitations with the zig-zag magnetic background.
The experiment shows that the partial melting of the short-ranged zig-zag magnetic background is mapped into the rearrangement of the narrow $E_{1g}^{(1)}\pm E_{2u}^{(2)}$ QMOs, as expected on the basis of the DFT calculations.

The time- and frequency-domain data reported in Ref.~\cite{Nembrini2016}, combined with single-color pump-probe measurements \cite{Alpichshev2015,Hinton2015}, provide a comprehensive picture of the relaxation dynamics in Na$_{2}$IrO$_{3}$ (see Fig.~\ref{fig_iridates}). 
\begin{itemize}
\item[a)] The initial photo-excitation with 1.55 eV photons creates a highly non-thermal distribution of electron-hole excitations across the Mott gap. Whitin $\sim$200 fs, these excitations relax through electron-electron interactions and the coupling with optical phonons. This process leads to the fast accumulation of doublons (i.e. doubly occupied Ir sites) and holons (i.e. unoccupied Ir sites) at the bottom (top) of the UHB (LHB), whose recombination is prevented by the large value of the gap. 
\item[b)] As recently demonstrated in Refs.~\cite{Alpichshev2015} and \cite{Hinton2015}, the interaction with the local magnetic environment drives the binding of these non-thermal excitations into local excitons.
The effective doublon-holon interaction is mediated by the combined presence in the microscopic Hamiltonian of the Kitaev terms and of a weak Heisenberg-type ordering term and, consequently, grows with the distance and progressively trap them into pairs \cite{Alpichshev2015}. These constraints slow down the rate of energy exchange with the local magnetic background, which is completed only within few picoseconds.
\item[c)]
The localized photoexcited excitons degrade within few picoseconds via the delocalization on the Ir hexagons, thus recovering a quasi-molecular picture of excited charges immersed in a zig-zag background, characterized
by a correlation length of the order of 1.5-2 nm. As demonstrated in Ref.~\cite{Mazin2012}, such a short coherence length coincides with the length-scale necessary for a full tight-binding description of the Na$_{2}$IrO$_{3}$ band structure in terms of QMOs. Considering the almost full zig-zag polarization of the $E_{1g}^{(1)}\pm E_{2u}^{(2)}$ QMOs (see Fig. \ref{fig_NIO}), the excess charge excitations in the UHB are intrinsically associated to the weakening of the zig-zag order, thus driving an effective heating of the spin background. As a feedback, the perturbation of the zig-zag magnetic order leads to a renormalization of
the strongly coupled $E_{1g}^{(1)}\pm E_{2u}^{(2)}$ at $\approx$−1 eV binding energy, that allows mapping the low-energy magnetic dynamics onto the variation of the optical properties at 1.4-1.7 eV.
\end{itemize}

These non-equilibrium results have two fundamental consequences. First, they show that the validity of the fully-localized ($J_{eff}$) and quasi-delocalized (QMO) scenarios strongly depends on the energy scale considered: while the low-energy magnetic dynamics can be described within a picture of localized moments characterized by bonding-dependent (Kitaev) interactions, QMOs are the effective
building blocks of the physics at binding energies larger than $\approx$1 eV. Second, they demonstrate that the local demagnetization processes can be mapped into specific optical transitions in the near-infrared/visible energy range by exploiting the intertwining between the magnetic and orbital degrees of freedom (QMO). This finding provides a new tool for investigating and controlling the ultrafast magnetic dynamics and, consequently, for achieving a better understanding of the
equilibrium properties of the system.

\subsection{Square-lattice strontium iridate}
In Sr$_{2}$IrO$_{4}$, the corner-sharing IrO$_{6}$ octahedra form IrO$_{2}$ planes, perpendicular to the crystallographic $c$-axis, where the Ir ions are arranged in a square lattice. Because of a staggered
in-layer rotation of the octahedra, the unit cell, which coincides with the magnetic unit cell, is formed by four IrO$_{2}$ layers \cite{Crawford1994,Kim2008,Kim2009,Kim2012}.
Similarly to Na$_{2}$IrO$_{3}$, $5$ electrons occupy the $t_{2g}$ manifold induced by the octahedral crystal-field. Because of the simpler topology of the square lattice, which does not induce neither additional magnetic frustration
nor privileged hopping paths, the localized $J_{eff}$ picture is solid and widely accepted \cite{Kim2008}: the SOC splits the six $t_{2g}$ levels and returns two half-filled $J_{eff}=1/2$ levels and four completely-filled
$J_{eff}=3/2$ levels, further split in two doublets by a residual tetragonal (distortion) crystal field \cite{Kim2014}. According to this scheme, the system behaves as a robust Mott insulator once the
effect of the modest on-site Coulomb repulsion $U$ is properly taken into account \cite{Kim2008}. The charge gap, estimated by optical conductivity \cite{Moon2006} and resistivity \cite{Kim2012} measurements is $\sim$100
meV. The Sr$_{2}$IrO$_{4}$ magnetic properties can be efficiently described by a Heisenberg Hamiltonian \cite{Kim2012,Kim2014} with exchange terms up to third neighbors in plane ($J_{\mathrm{h}}=60$,
$J_\mathrm{hh}=-20$, and $J_\mathrm{hhh}=15$ meV) \cite{Kim2012}, while the out-of-plane
term is as low as $J_{\perp}\approx$1 $\mu$eV. The Heisenberg model accounts for the canted in-plane
anti-ferromagnetic ordering with a critical N\'{e}el temperature of about 240 K \cite{Cao1998}.

\subsubsection*{\textbf{Direct observation of the ultrafast dynamics of magnetic correlations}.}
Electronic and magnetic correlations play
a critical role in the formation of transient non-thermal magnetic states, but an appropriate
tool for probing the momentum and energy dependence of these correlations has been lacking
so far. In Ref.~\cite{Dean2016} the first time-resolved magnetic Resonant Inelastic X-ray Scattering (tr-RIXS)
experiment was performed to directly probe the magnetization dynamics in Sr$_{2}$IrO$_{4}$, after
the impulsive photo-doping across the Mott gap. X-ray photons, produced by a free electron laser and
tuned to the Ir $L_{3}$ resonance in order to couple with the spin
degree of freedom via the resonant magnetic X-ray scattering mechanism, are used as probe.
The possibility of measuring the momentum transfer $Q$, the energy loss $E_{l}$, and the time delay between the pump and the probe, allows to fully characterize the transient magnetic response of the system. The experiment was performed
at 110 K, well below the N\'{e}el ordering temperature. The $(-3,\:-2,\:28)$ magnetic Bragg peak intensity, which is sensitive to the presence of 3D magnetic order, has been measured as a function of the time delay and
the pump fluence. For fluences larger than 5 mJ/cm$^{2}$, corresponding to the excitation of a substantial fraction of the sites within the illuminated volume, the residual intensity of the peak falls below $10\%$ indicating the partial melting of the 3D order. Then, tuning the pump fluence at 6 mJ/cm$^{2}$, the RIXS energy loss spectra has
been measured after the photo-excitation and compared to the unperturbed
one. The spectra exhibit two main features: (a) a dispersing magnon and (b) an orbital excitation
at $\approx$600 meV corresponding to the excitation of a hole from the $J_{eff}=1/2$ state to the $J_{eff}=3/2$ one \cite{Ishii2011,Kim2012,Kim2014}.
The intensity of the orbital excitation, which is proportional to the population
of the two $J_{eff}$ states, is very similar before and after (t=2 ps) the
photo-excitation, thus giving an upper limit to the lifetime of the $J_{eff}=3/2$
photo-excited carriers. On the other hand, the transient optical reflectivity
measured at 1.55 eV (800 nm) clearly shows that an appreciable population
of photo-excited carriers persists at 6 mJ/cm$^{2}$ and 2 ps time
delay. From these two combined measurements, the authors argue that 2 ps is also the upper limit of
the lifetime of electron-hole pairs, or doublons, living in the upper
and lower Hubbard bands \cite{Dean2016}. 

The magnon dispersion measured by RIXS perfectly
agrees with the equilibrium excitation spectrum modeled via the Heisenberg model \cite{Kim2012}:
the minimum (zero) resides at $Q$=$\left(\pi,\pi\right)$, the maximum
($\sim$200 meV) at $Q$=$\left(\pi,0\right)$. Given the extremely small
inter-layer coupling, the ideally 3D experimental dispersion essentially
resembles the theoretical 2D one at equilibrium \cite{Kim2012}. Despite
the collapse of the 3D magnetic order, magnons are observed in the
RIXS spectra at both $Q$ points in the transient state after the
pump with a dispersion practically identical to the equilibrium one
at $Q$=$\left(\pi,0\right)$ and slightly different at $\left(\pi,\pi\right)$.
These findings show that: i) the 2D correlations retain their N\'{e}el
nature in the transient state up to 2 ps after the pump; ii) the melting
of the 3D state is related to the destruction of the inter-layer correlations.
The differences at $Q$=$\left(\pi,\pi\right)$, which recovers only on
a much longer time scale ($100$ to over $1000$ ps), can
suggest the difficulty of recovering low-energy excitations with respect
to high-energy ones (e.g. those at $\sim$200 meV and $Q$=$\left(\pi,0\right)$).
Moreover, the increase of the low energy spectral intensity at $Q$=$\left(\pi,\pi\right)$
and the related depletion of the one at $\sim$100 meV cannot be explained
by simple thermal heating, which would lead to a uniform relative
broadening, and the residual photo-excited carriers in the transient
state could be held responsible for them. 

These results  demonstrate that the photo-excitation
of carriers across the Mott gap can be used to create a non-thermal magnetic state in which the equilibrium 3D magnetic correlations have been melted (in less than $\sim$0.3 ps), while non-thermal 2D correlations transiently survive for a longer time (10-100 ps). The complete recovery of the 3D magnetic state requires 100-1000 ps (in agreement with the ratio $J/J_{\perp}$). The creation of the \textit{artificial} magnetic state with 2D correlations is crucially connected to the presence of long-lived charge excitations in the Hubbard bands.This finding opens the possibility of developing selective charge-excitation protocols to transiently control the magnetic state in strontium iridates.

\subsection{Towards the ultrafast magnetic control in 5d materials}
The two paradigmatic examples of the honeycomb-lattice Na$_2$IrO$_3$ and the square-lattice Sr$_2$IrO$_3$ demonstrate that the creation of non-thermal electronic populations can be exploited to photoinduce transient magnetic states, which are potentially tunable and reversible. The nature of these transient states crucially depends on the topology of the lattice, on the strength and directionality of the short-range magnetic interactions and on the lifetime and orbital character of the charge excitations. Here we summarize the key concepts that can be exploited for developing novel schemes for the ultrafast manipulation of magnetic states in $5d$ transition-metal oxides:
\begin{itemize}
\item The interplay of SOC, correlations and topology of the lattice, typical
of $5d$ TMOs and relativistic correlated materials in general, allows
to map the local demagnetization processes into specific optical transitions
in the near-infrared/visible energy range. The other way around, the resonant photostimulation of high energy charge states with specific orbital components can be exploited to selectively trigger magnetic excitations.
\item The competition between different magnetic configurations $-$ e.g. zig-zag and stripy antiferromagnetic phases $-$, typical of frustrated correlated magnets, can be exploited
for achieving a controlled magnetic switching via the selective excitation
of charge states coupled to magnetic correlations with different net magnetization in terms of spatial
arrangement and polarization.
\end{itemize}

In the near future, we foresee that the use and development of ultrafast techniques
will become more and more fundamental to understand how the magnetic dynamics of correlated 5$d$ materials can be functionalized and manipulated by light. The combination of ultrafast photo-doping with the chemical surface \cite{Kim2016} doping will provide new opportunities to discover additional emergent properties of scientific and technological interest.

\section{Coherent electronics in few-atomic-layers heterostructures}

Controlling the electronic quantum coherence in solids at ambient conditions is a long sought-after target in condensed matter physics. Quantum pathways could be exploited to coherently convert photons into charge excitations, thus tremendously increasing the performances of optoelectronic, spintronic and photovoltaic devices. The possibility of  quantum-coherently guiding and controlling electronic excitations would provide the building block to develop robust quantum systems that could be manipulated at ambient conditions, with impact on quantum-information processes and communication technologies \cite{Awschalom2007}.
Besides the technological impact, the loss of quantum coherence in solids is usually the primary relaxation mechanism and is mainly related to quasi-elastic scattering processes. As a consequence, the direct access to the decoherence timescales in materials is expected to provide a new platform to investigate some of the most fascinating and unexplored problems of condensed matter physics, such as:   
\begin{itemize}
\item[-]
The hierarchy of the interactions with the magnetic, orbital and lattice degrees of freedom in leading the dephasing and the eventual thermalization of many-body excitations in TMOs and other correlated materials.
\item[-]
The way quantum-coherent wavefunctions evolve, on a longer timescale, into incoherent charge, spin and lattice excitations, whose propagation is regulated by \textit{incoherent} transport laws.
\item[-]
The possibility of manipulating the electronic properties of materials on the verge of correlation-driven phase transitions by creating suitable coherent superposition of quantum states that evolve faster than the decoherence processes.
\item[-]
The conversion of optical excitations into electron-hole pairs, which can be collected along quantum-coherent pathways for photon-harvesting applications in engineered heterostructures.
\end{itemize}

Unfortunately, the quantum nature of electronic excitations in solids at liquid nitrogen or ambient temperatures is usually washed out on the femtosecond timescale by the interaction with the incoherent fluctuations of the charge, phonon and spin backgrounds (thermal bath). The most common strategy to overcome this problem is to \textit{freeze} the incoherent thermal bath by cooling down to extremely low temperatures ($<$2 K) and investigating, for example, the coherent effects on the steady-state transport of charges or optical excitons in semiconductor nano- or meso-structures \cite{Beenakker1991}. 

An alternative approach consists in directly accessing and, eventually, controlling the dephasing processes at high temperatures on their relevant timescales. However, the attempts to tune and exploit the electronic quantum coherence in correlated nanomaterials have so far been hindered by the lack of fundamental knowledge $-$ both theoretical and experimental $-$ of the mechanisms driving the ultrafast dephasing. In this section we will discuss and review the concepts and the most recent results that are expected to lay the foundation of this field.

\subsection{The decoherence timescales in correlated materials}
\label{sec_dectimescale}
In solids, the simplest example of coherent transport is given by the Bloch oscillations. When electrons, described by Bloch wavefunctions, are free from any scattering process, they respond to a steady electric field $E$ in an oscillatory way, with period T$_B$=2$\pi\hbar$/$aeE$ ($a$ is the lattice constant, $e$ the electronic charge) \cite{Kittel}. Even though in bulk materials this period is much larger than the dephasing time, thus making oscillations vary hard to observe, the recent advances in the atomic-scale synthesis of oxide heterostructures \cite{Hwang2012} have opened very intriguing scenarios. For example, the polar nature of many TMO heterostructures, such as LaAlO$_3$/SrTiO$_3$ and LaVO$_3$/SrTiO$_3$, gives rise to intrinsic electric fields \cite{Assmann2013} up to $E_{in}$=0.08 eV/\AA. In these systems, the corresponding period of the Bloch oscillations is T$_B\sim$12 fs, which is very close to both the expected dephasing time and the time necessary to eventually collect the charges across few atomic monolayers. This observation suggests the possibility of exploiting quantum-coherent pathways to enhance the conversion efficiency in suitable engineered few-monolayers systems, provided the excitations are collected faster than the time necessary to completely loose the initial quantum-coherence.   

\subsubsection*{\textbf{Decoherence and the density matrix}} In order to properly introduce the concept of dephasing of a generic state quantum state $\vert\psi\rangle$, it is necessary to make use of the density matrix, $\widehat{\rho}$(t)=$\vert\psi\rangle\langle\psi\vert$. For sake of simplicity, we limit the discussion to a two-level system, that accounts for the ground ($\vert\psi_0\rangle$) and excited ($\vert\psi_e\rangle$) states. Any form of quantum-coherent excitation prepares the system into a superposition of the ground and excited states: $\vert\psi\rangle$=$a_0$(t)$\vert\psi_0\rangle$+$a_e$(t)$\vert\psi_e\rangle$. Assuming that $a_{0(e)}$(t)=$c_{0(e)}$(t)exp(-$i\omega_{0(e)}$t), where $c_{0(e)}$(t) are slowly-varying complex functions of time and $\omega_{0(e)}$=$E_{0(e)}$/$\hbar$ are the eigenfrequencies of the ground and excited states, the matrix representation of $\widehat{\rho}$ reads:
\begin{equation}
\widehat{\rho}\mathrm{(t)}=\left[\begin{array}{cc}
\rho_{00}&\rho_{0e}\\
\rho_{e0}&\rho_{ee}\\
\end{array}\right]=\left[\begin{array}{cc}
|c_0\mathrm{(t)}|^2&c_0\mathrm{(t)}c^*_e\mathrm{(t)}e^{-i\omega_{0e}\mathrm{t}}\\
\\
c_e\mathrm{(t)}c^*_0\mathrm{(t)}e^{i\omega_{0e}\mathrm{t}}&|c_e\mathrm{(t)}|^2\\
\end{array}
\right] \label{definition_of_DM}
\end{equation}
where $\omega_{0e}$=($E_0$-$E_e$)/$\hbar$.
In this form, the diagonal terms describe the time-dependent populations of the ground and excited states, while the off-diagonal ones define the coherences of the system, which can be related to the macroscopic polarization $\overrightarrow{P}$(t)=$N/V\overrightarrow{\mu}_{0e}\left(\rho_{0e}+\rho_{e0}\right)$, via the atomic dipole moments $\overrightarrow{\mu}_{0e}$=$\langle\psi_0\vert \overrightarrow{\widehat{\mu}} \vert\psi_e\rangle$ with density $N/V$. Importantly, the density matrix maintains exactly the same physical meaning in the case of statistical ensembles, where the matrix elements should be calculated as the averages over the population $s$: $\overline{\rho}_{0e}$=1/$N\sum_s{\rho}^{(s)}_{0e}$. Even though the microscopic treatment of the time-evolution of $\widehat{\rho}$ strongly depends on the specific system considered and goes beyond the scope of the present work, the decoherence processes can be accounted for in a phenomenological way by introducing a decoherence timescale, $\tau_D$, that depends on the interaction with the environment and produces the exponential decay of the off-diagonal matrix elements:
\begin{equation}
\widehat{\overline{\rho}}\mathrm{(t)}=\left[\begin{array}{cc}
\overline{|c_0\mathrm{(t)}|}^2&\overline{c_0c^*_e}e^{-i\omega_{0e}\mathrm{t}}e^{-\mathrm{t}/\tau_D}\\
\\
\overline{c_ec^*_0}e^{i\omega_{0e}\mathrm{t}}e^{-\mathrm{t}/\tau_D}&\overline{|c_e\mathrm{(t)}|}^2
\end{array}
\right] \label{definition_of_DM}
\end{equation}
$\tau_D$ thus characterizes the evolution of a macroscopically quantum-coherent state into an incoherent statistical ensemble ($\rho_{0e}, \rho_{e0}\rightarrow$0) and is determined by all the quasi-elastic primary relaxation processes that destroy the macroscopic coherence without altering the population of the levels.
The precise knowledge of $\tau_D$ in solids represents the prerequisite for the design of experiments and theories aimed at measuring and describing the decoherence processes and the transport properties on the timescales shorter than those necessary to achieve the incoherent transport regime.

\subsubsection*{\textbf{Excitonic resonances in correlated materials}}
\begin{figure}
\begin{center}
\includegraphics[width=1.1\textwidth]{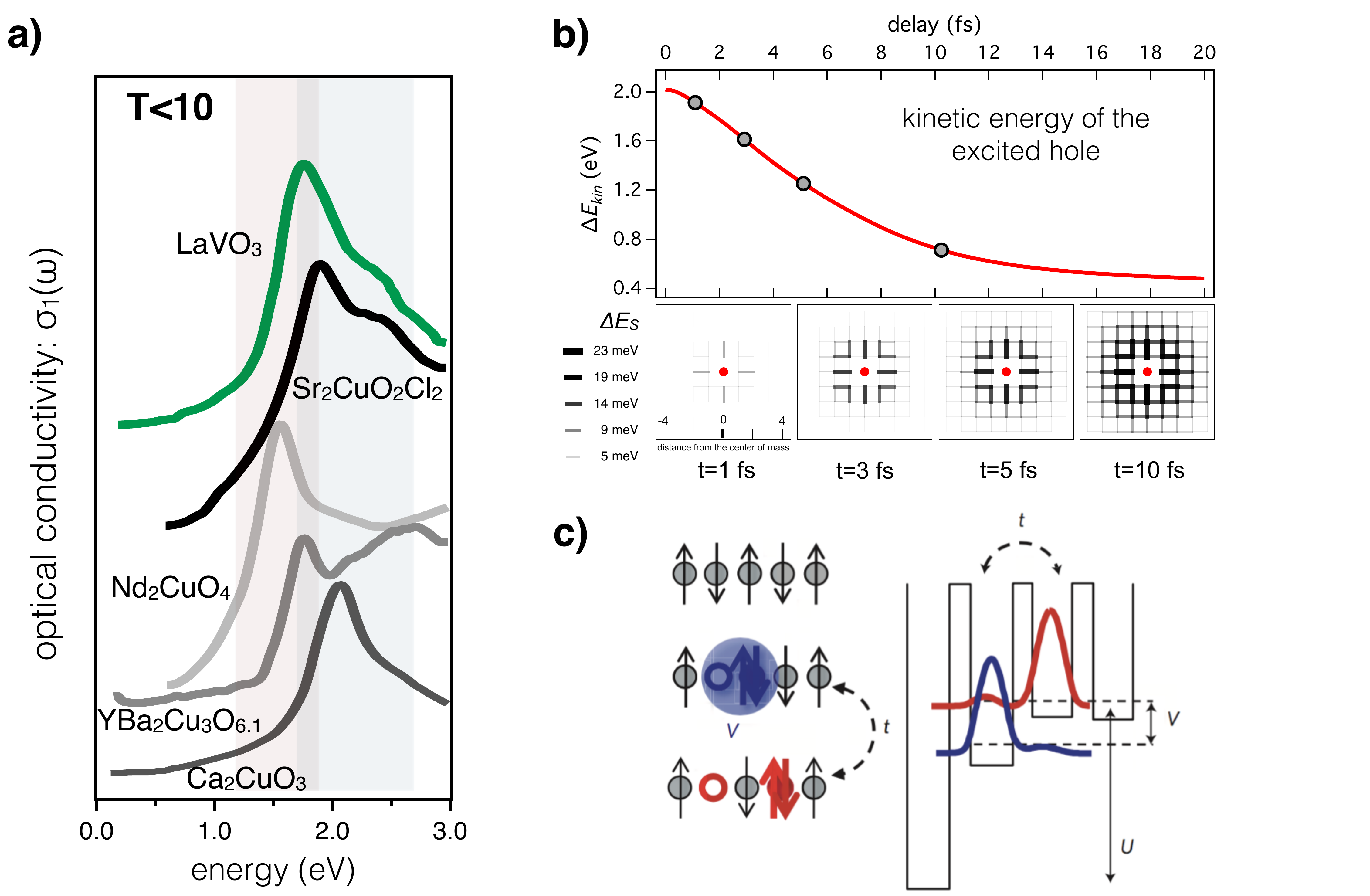} 
\caption{a) Low-temperature charge transfer and exciton resonances for different families of transition metal oxides \cite{Uchida1991,Miyasaka2002}. The colored areas represent the energy regions that can be accessed by sub-10 fs NOPA pumped by amplified ultrafast lasers for high-resolution 2D spectroscopy (see Section \ref{2Dspec}). Taken from Ref. \cite{Giannetti2016b}. b) Time evolution of the hole kinetic energy variation within the $t$-$J$ model. The bottom panels display the time evolution of the relative energy increase of the local antiferromagnetic bonds on a 9x9 square lattice around the photo-excited hole (red circle). The thickness of the black segments is proportional to the energy stored in each bond. Taken from \cite{DalConte2015}. c) Cartoon of the ultrafast dynamics of the excited holon–doublon bound states and of the ionized holon–doublon states. The quantum-interference between bound and ionized holon–doublon states can be mapped onto that of two potential-energy wells offset by $V$ and coupled by tunnelling amplitude $t$. Taken from Ref. \cite{Wall2010}.}
\label{fig_CT}
\end{center}
\end{figure}

In order to observe coherent phenomena in solids, it is necessary to find well defined excitonic transitions, whose dephasing time can be accessed by ultrafast techniques. While the study of low-temperature excitons in semiconductors is a well established field \cite{Cundiff2013}, correlated materials and TMOs constitute an almost unexplored world, though very promising. Let us consider the antiferromagnetic Mott (or charge-transfer) insulators, such as the copper oxides parent compounds. The lowest excitation is the charge-transfer (CT) of a localized Cu-3$d_{x^2−y^2}$ hole to its neighbouring O-2$p_{x,y}$ orbitals, with an energy cost $\Delta_{CT}\simeq$1.5-2 eV. In the optical conductivity, this process is revealed by a typical charge-transfer (CT) edge, which defines the onset of optical absorption by particle-hole excitations. However, the local CT excitation is a many-body process and can be \textit{dressed} by structural deformations or rearrangements of the spin-background (lattice- or spin-polarons), or even form a bound state (Hubbard exciton), thus forming a local exciton which can hop in the lattice and in the ordered antiferromagnetic background. This physics is entirely contained in the ubiquitous structures that are observed at the CT edge of TMOs  (see Fig. \ref{fig_CT}) and whose nature is still subject of intense debate \cite{Tokura1990,Uchida1991,Miyasaka2002,Gossling2008,Reul2012}. From the linewidth of these resonances, which ranges between 0.2 and 0.4 eV, it is possible to estimate the upper bound of the excitonic lifetimes (1.5-3 fs). Although very short, these values are close to the current limit of ultrafast technologies that pushed the temporal resolution well below 10 fs \cite{Brida2010,Brida2014,Nillon2014}. However, the total linewidths in TMOs are usually severely affected by the \textit{inhomogeneous} broadening related to the presence of defects or intrinsic electronic/lattice inhomogeneities typical of correlated materials \cite{Fradkin2012}. The real dephasing time, which corresponds to the inverse of the \textit{homogenous} contribution to the total linewidth, should be evaluated either by starting from microscopic theories or by developing suitable multi-dimensional ultrafast techniques, as discussed in Sec. \ref{2Dspec}.
We stress that similar physics is present in many correlated TMOs, such as vanadium oxides (LaVO$_3$, V$_2$O$_3$) and Ir-based relativistic Mott insulators (Sr$_2$IrO$_4$, Na$_2$IrO$_3$). The possibility of investigating the decoherence of excitonic resonances at the edge of the Mott (charge-transfer) gap is thus very general to many classes of TMOs that can be exploited for applications.

\subsubsection*{\textbf{Time-domain models}} 
In conventional materials, the decoherence of excitonic resonances is mainly led by the direct exciton-exciton interactions, which can be limited by properly tuning the intensity of the light excitation, and by the scattering with bosonic fluctuations, such as the phonons, that is effective on the timescale $\hbar$/$\Omega_B$, $\Omega_B$ being the typical boson frequency. When considering copper oxides, the strong on-site Coulomb repulsion $U$ between two electrons occupying the same Cu-3$d_{x^2-y^2}$ orbital introduces a new dynamics characterized by the energy scale $J$=2$t_{\mathrm{h}}^2/U\simeq$120 meV, where $t_{\mathrm{h}}$ is the nearest-neighbor hopping. The antiferromagnetic coupling $J$ between neighboring sites arises due to the virtual hopping of holes into already occupied sites through Anderson's superexchange mechanism. While the phonon cut-off in cuprates is about 70 meV, corresponding to a timescales of $\sim$10 fs, the spectrum of the spin fluctuations extends up to $2J\sim$250 meV ($\hbar$/2$J\simeq$3 fs) and it is believed to constitute the primary scattering channel for the charge excitations photoinduced in copper oxides by the ultrashort pump pulses \cite{Giannetti2016}.

In the less favorable scenario, the quantum coherence of the excitonic resonances at the gap edge in Mott and charge-transfer insulators is lost at each hopping process. If we assume a typical hopping integral of 0.1-0.3 eV for TMOs, we can estimate a decoherence time in the order of 2-6 fs ($\tau_D$=$\hbar$/$t_{\mathrm{h}}$, $\hbar$=658 meVfs). However, there are many effects that can concur to increase this value: i) the formation of “dressed” coherent quasiparticles is expected to increase the effective mass, thus reducing their mobility within the lattice \cite{Kane1989,Lee2006}; ii) in multi-band systems, the possibility of hopping on the oxygen sub-lattice strongly reduces the effect of spin-fluctuations on the electron dynamics \cite{Ebrahimnejad2014}. 

The ultrafast relaxation of Hubbard excitons (HE) in Mott insulators has been directly tackled by studying a simplified version of the single-band Hubbard model. A generalized $t$-$J$ model for insulators is obtained by performing a canonical transformation of the Hubbard model that, at the lowest order in $t$/$U$, decouples sectors with different numbers of HEs, thus accounting for the recombination of HEs \cite{Lenarcic2013,Lenarcic2014}. Within this model, the decay dynamics is given by a two-step process: i) the formation of $s$-type HEs that are trapped at the bottom of the Mott gap; ii) the decay of the bound holon-doublon exciton via multimagnon emission, which is rather effective as a consequence of the strong charge-spin coupling in 2D systems. Using realistic values of $J$ and $t_{\mathrm{h}}$ for cuprates, the estimated lifetime of HEs ranges from few tens to hundreds femtoseconds \cite{Lenarcic2013}. 
As a further step in reproducing the ultrafast dynamics in correlated insulators, the electron-phonon interaction can be accounted for by the Holstein model that includes the coupling with a dispersionless phonon mode. The relaxation in the full Hubbard-Holstein model driven out of equilibrium by an ultrashort light pulse has been recently performed on eight lattice sites \cite{DeFilippis2012}. At non-zero electron-phonon interaction, signature of the coherent evolution of the charge, spin and phonon populations are found to persist up to hundreds of femtoseconds.

The problem of the lifetime of photo-excited charge carriers in doped materials is even more difficult. At non-zero doping, strongly renormalized quasiparticles states at the Fermi level coexist with incoherent excitations at the energy scale of $\Delta_{CT}$. The complete treatment of the problem requires solving the full Hubbard model out of equilibrium, which goes beyond the present capabilities. However, a suitable approximations is to study the dynamics of a single hole within the $t$-$J$ model. The photoexcitation process is accounted for by suddenly rising the kinetic energy of the hole and letting it evolve in the spin-lattice background \cite{Golez2012,Kogoj2014,Golez2014,DalConte2015} (see Fig. \ref{fig_CT}b). In general, the scattering with magnetic fluctuations is found to be the fastest relaxation process. The reason of the efficiency of the charge-spin scattering is that almost any hopping process leads to the perturbation of the antiferromagnetic spin background. In a square lattice, the lifetime of excited carriers corresponds to a renormalized hopping time \cite{Golez2014,DalConte2015}: $\tau_D$=($\hbar$/$t_{\mathrm{h}}$)($J$/$t_{\mathrm{h}}$)$^{-2/3}$. When considering the typical values of $J \simeq$120 meV and $t_{\mathrm{h}} \simeq$360 meV for copper oxides, a relaxation time $\tau_D \simeq$4 fs is estimated. 

This estimation of the timescale for the coupling with antiferromagnetic spin fluctuations in copper oxides has been also confirmed by the solution of the full Hubbard model with a non-equilibrium version of the dynamical cluster approximation \cite{Eckstein2014}. In the Mott insulators, the relaxation rate of the high-energy photo-excited charge carriers is found to be $\simeq$10-20 fs. The relaxation process becomes more complex at finite doping, where the direct charge-charge interactions open additional scattering channels that lead to a faster relaxation of the photo-excited holes (electrons).

\subsubsection*{\textbf{Time-domain experiments}} 
The experimental study of the decoherence processes in correlated materials has been mainly hindered by the lack of suitable techniques to directly access the off-diagonal elements of $\widehat{\overline{\rho}}\mathrm{(t)}$, with a temporal resolution of the order of $\sim$10 fs. Nonetheless, the use of broadband sub-10 fs pulses recently allowed to perform time-resolved spectroscopic measurements that provided important insights.

The decoherence of the Hubbard excitons has been observed in the organic salt bis(ethylenedithio)tetrathiafulvalene-difluorotetracyanoquinodimethane (ET-F$_2$TCNQ) by performing broadband pump-probe experiments with $\simeq$9 fs pulses with a spectrum covering the 0.55-1 eV energy range. After the  photoexcitation, the quantum interference between the HEs and ionized holon-doublon pairs (see Fig. \ref{fig_CT}c), that are created when the holons and doublons move far apart in the 1D lattice, has been observed under the form of an oscillation at 25 THz (period of $\simeq$40 fs) \cite{Wall2010}. These results confirm that, in Mott (charge-transfer) insulators, the photoexcited HEs maintain the initial quantum coherence on a timescale of several tens of femtoseconds.

In doped Mott insulators, such as doped cuprates, the timescale related to the coupling with short-range spin fluctuations has been recently inferred by broadband pump-probe measuerements in different families of hole-doped copper oxides \cite{DalConte2015}. The temporal resolution of $<$10 fs, that is of the order of the timescale (1/2$J$) of the direct coupling to magnetic fluctuations, disclosed the possibility of directly observing the energy exchange between the initially photoexcited charge carriers and the bosonic bath, constituted by the phonons and magnetic fluctuations. The measured timescale, $\tau_D \simeq$16 fs, has been benchmarked against numerical calculations based on the $t_{\mathrm{h}}$/$J$ model, in which the relaxation of the photo-excited charges is achieved via inelastic scattering with short-range antiferromagnetic excitations \cite{DalConte2015}.

\subsection{The “transport problem” in oxide heterostructures}
\label{sec_transport_hetero}

\begin{figure}
\begin{center}
\includegraphics[width=1\textwidth]{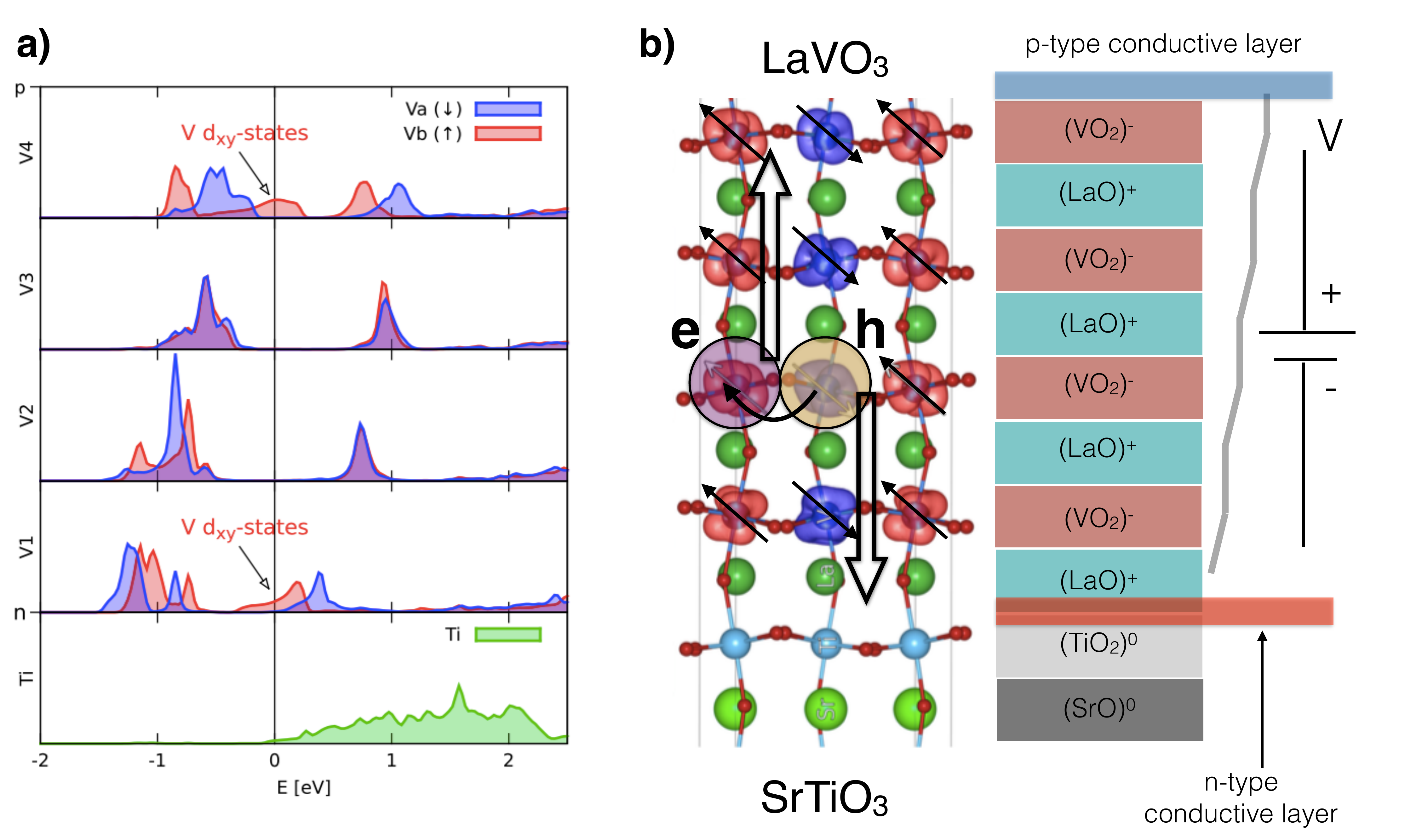} 
\caption{a) Layer-by-layer density of states (DOS) of the LaVO$_3$/SrTiO$_3$ heterostructure as calculated by DFT+$U$ \cite{Assmann2013}. The blue and red colors indicate the DOS for each of the two inequivalent (opposite spin) V sites per layer. b) Generation and collection of optical e-h excitations in the LaVO$_3$/SrTiO$_3$ heterostructure. Partially taken from \cite{Assmann2013}.}
\label{fig_LVO}
\end{center}
\end{figure}
As discussed in the previous section, the typical decoherence timescale for electron-hole excitations at ambient temperature in TMOs can range between 4 fs and several tens of femtoseconds, depending on the specific characteristics (charge carrier density, hopping $t_{\mathrm{h}}$) of the system considered. The recent advances in ultrafast science made these timescales directly achievable, thus providing the platform for the study of coherent-transport phenomena in correlated materials. 

From the technological standpoint, the possibility of tailoring the coherent dynamics on ultrafast timescales would have a dramatic impact on the working principle of few-atomic-layers oxide-based devices. To properly make the case, we start from a paradigmatic example, i.e. LaVO$_3$/SrTiO$_3$ (LVO/STO) heterostructures (see Figure \ref{fig_LVO}), that have been recently suggested for photovoltaic applications \cite{Assmann2013}. At 300 K, LVO is a orthorhombic paramagnetic Mott insulator (direct gap of $\simeq$1.1 eV) with 2 electrons in the crystal-field split 3$d$-$t_{2g}$ orbitals. The fundamental optical excitation is the transition from the lower (LHB) to the upper Hubbard band (UHB), which corresponds to moving an electron from one V atom to its neighbouring V-site. At 140 K, LVO transforms into a monoclinic structure with an antiferromagnetic $C$-type order ($c$-axis ferromagnetic stacking of antiferromagnetic $ab$ layers). Interestingly, the orthorhombic LVO is one of the few Mott insulators which exhibits large quantum fluctuations at room temperature \cite{Raychaudhury2007}, in the sense that the orbital occupation of the $t_{2g}$ orbitals is strongly fluctuating on a frequency range which is determined by quantum effects instead of $k_BT$. However, these strong incoherent orbital fluctuations are almost completely suppressed in the monoclinic 140 K phase, when they turn into a phase-coherent condensate (orbital ordering) in which the orbital occupation assumes a well defined (phase stable) pattern in space and time. 
The exploitation of the LVO properties is based on the possibility of creating heterostructures in which the chemical/structural composition of each layer is controlled at the atomic scale \cite{Hwang2012}. Let us consider the multilayer structure constituted by 4 LVO unit cells interfaced to SrTiO$_3$, as suggested in Ref. \cite{Assmann2013}. Density Functional Theory (DFT) calculations (see Fig. \ref{fig_LVO}) show the layer-by-layer density density of states (DOS) of the V-3$d_{xy}$ states. Interestingly, the shift of the bands between neighboring layers indicates the formation of a potential gradient between the (LaO)$^+$-(VO$_2$)$^-$ planes, that is estimated as $E_{in}$=0.08 eV/\AA. Furthermore, the intrinsic presence of conducting states (of $d_{xy}$ character) at the LVO/STO interfaces provides a natural way of collecting the electron-hole excitations generated within the LVO layers. The working principle of the LVO/STO heterostructure can be described by the following steps: i) the absorption of an above-gap photon creates a many-body excitation $\vert\psi_e(\textbf{r})\rangle$ that consists in a local doublon (doubly occupied V-3$d_{xy}$ orbital) and holon (unoccupied V-3$d_{xy}$ orbital) pair; ii) the charged doublons and holons, subject to the intrinsic $E_{in}$, hop across the few LVO atomic layers and are collected by the two opposite interface conductive layers thus generating a photocurrent.  

The efficiency of such a device is crucially based on the pathways available to collect the initial electron-hole excitons. Given that the exciton recombination time ($>$100 fs, see Ref. \cite{Lenarcic2014}) is large as compared to the time necessary to collect the charges, the quasi elastic scattering processes, which dephase the initial wavefunction without changing the population, are expected to play a crucial role. While a large scattering rate would be detrimental for the collection of the charges, the absence of any scattering process would lead to Bloch oscillations without efficiently establishing any photocurrent. This naive expectation has been recently demonstrated by non-equilibrium dynamical mean-field theory (DMFT) simulation of the transport properties of a multilayer Mott insulator in which doublons and holes are created by photoexcitation \cite{Eckstein2013}. 

The natural question emerging from these considerations is whether the timescale necessary to collect the charges is compatible with the emergence of transport phenomena along \textbf{quantum walks}, which can be faster and more efficient than classical ones \cite{Mohseni2008,Ishizaki2009}.  
Considering the hopping terms estimated by Local Density Approximation (LDA) calculations combined with DMFT \cite{Raychaudhury2007}, we note that the most efficient channel for collecting charges along the $c$-axis (perpendicular to the (VO$_2$)$^-$ planes) in the orthorombic phase is mainly related to the hopping between adjacent 3$d_{xz}$ orbitals, with a hopping integral $t_{xz,xz}\simeq$190 meV. In the low-temperature monoclinic phase ($T<$140 K), the most favorable hopping is given by the overlap between 3$d_{xz}$ and 3$d_{yz}$ orbitals with an hopping integral $t_{xz,yz}\simeq$150 meV. Assuming that the photo-exciton is created in one of the internal LVO layers, the time necessary to collect the charges ($\tau_{coll}$) is given at the most by two times the inverse hopping integral, i.e. $\tau_{coll}$=6-8 fs. This value is very close to the $\tau_D$ estimated in Section \ref{sec_dectimescale}, thus suggesting that the quantum laws may greatly affect the transport properties of nanoscaled TMO devices, such as LVO/STO heterostructures.

\subsubsection*{\textbf{Coherent transport in bio-systems}}
In the simplest quantum-transport models, complex interacting systems are modelled as a quantum network of two-level systems (sites), mutually coupled via hopping amplitude and interacting with both a thermal bath and a continuum of states (the collector). The \textit{incoherent} hopping motion of the charge excitation corresponds to the case in which $\tau_D$ is shorter than the hopping time and only the population of the multi-levels, i.e., the diagonal terms of the density matrix, contributes to the transport. Quantum-coherent effects can emerge, for example, when $\tau_D$ is much larger than the typical hopping time and, as a consequence, the off-diagonal terms of the density matrix can give rise to novel effects mainly related to the interference of the wavefunctions describing each site.
Considering the case of photo-stimulated excitons propagating in a quantum network coupled to the environment, many concepts have been introduced to describe the way local excitations are generated, transported and collected in photosynthetic light harvesting complexes \cite{Scholes2011}. The study of coherent effects in bio-systems at physiological temperatures and the role of the coupling with the environment was largely boosted by the discovery that quantum coherence is preserved in Fenna-Matthews-Olson (FMO) bacteriochlorophyll complexes \cite{Engel2007} and in marine algae \cite{Collini2010} on timescales of the order of hundreds femtoseconds. As a consequence, the first stage of the energy transfer within the light harvesting complexes can be dominated by wave-like phenomena,  that possibly enhance the collection efficiency by allowing the simultaneous probing of large areas of phase space to find the most efficient path.

In quantum-coherent models, the transport efficiency is strongly affected by the interaction with different environments, such as the coupling with the collector and the thermal bath. In general, the problem is treated by suitable master equations \cite{Ishizaki2009} that describe the time evolution of both the diagonal ($\rho_{ii}$) and off-diagonal ($\rho_{ij}$, with $i\neq j$) terms of the density matrix, where $i$($j$) labels the excited site. One of the simplest master equation which can be used to model the effects of different environments on excitonic transport is the Haken-Strobl master equation \cite{Celardo2012}:

\begin{equation}
\label{eq_HakenStrobl}
\frac{\partial}{\partial\mathrm{t}}\rho_{ij}(\mathrm{t})=-\frac{i}{\hbar}[H_{eff}\rho-\rho H^\dagger_{eff}]_{ij}-(1-\delta_{ij})\frac{\rho_{ij}}{\tau_D}
\end{equation}
where $H_{eff}$ is an effective non-Hermitian Hamiltonian which describes the interaction with different dissipative environments, such as the collector where the excitation can be trapped and the electromagnetic field where the excitation can be lost by recombination. The effective non-Hermitian Hamiltonian can be written as $H_{eff}$=$H_0$-$i\gamma_{ct} Q$. Here $H_0$ is the Hermitian Hamiltonian describing the closed system (which is the network of coupled two-level sistes) and also includes the static (time-independent) disorder induced by the environment; $Q$ describes the dissipation in the continuum channel (collector or electromagnetic field) with strength $\gamma_{ct}$.
The imaginary part  ($\Gamma_r$) of the $r$-th complex eigenvalue of $H_{eff}$  
represents the decay width of the correspondent eigenstate. The quantum dephasing (the decay of the off-diagonal matrix elements) is accounted for by the second right term of equation \ref{eq_HakenStrobl} and represents the effect of the external bath which the network is coupled to. Physically, the dephasing is induced by time-dependent fluctuations of the site energies with amplitude proportional to 1/$\tau_D$.

As the above simple model shows, quantum transport is affected by different parameters: the strength of the dephasing, the strength of static disorder and the strength of the coupling to the collector. Finding the optimal parameters is a difficult task since their interplay is highly non-trivial. Many different principles, such as the enhanced noise-assisted transport \cite{Rebentrost2009,Plenio2008}, the Goldilocks principle \cite{Lloyd2011}, and the Superradiance in transport \cite{Celardo2009,Celardo2012,Celardo2014} (opening-assisted quantum enhancement of transport), have been proposed in order to understand how optimal transport can be achieved as a function of these many different parameters. Here we summarize the main concepts that are expected to be of great relevance for the case of transport in heterostructures:

\begin{itemize}

\item \textbf{Superradiance} occurs when the energy levels of a quantum systems are coherently coupled to a common decay channel.
In short, it implies  the existence of states with a  cooperatively enhanced decay rate
and it has been shown to have important effects on transport efficiency in open systems: for example,
maximal efficiency in energy transport  in the FMO system is achieved in the vicinity of the superradiance
transition (ST).
Superradiance,  is usually reached
only above a critical coupling strength with the continuum (in the overlapping resonance regime):
$\Gamma/\Delta \geq 1 $
where $\Gamma=\langle\Gamma_r\rangle$ is the average decay width and $\Delta$ is the mean level spacing of the closed system.
Since  the superradiance transition is due to the coherent
constructive interference between the paths to the collector one might expect that any
consequences of ST would disappear in the presence of
dephasing and relaxation provided by the thermal bath. On the
contrary,  strong opening-assisted quantum enhancement of transport for the FMO
is seen  also at room temperature where the quantum transfer is up to two times faster than that obtained by an incoherent calculation \cite{Zhang2016}.
The existence of \textit{superradiant states}, that are mainly localized in the vicinity of the collectors, is intrinsically accompanied by that of \textit{subradiant} states which are long-lived and effectively decoupled from the environment. The existence of subradiant solutions opens the possibility of designing protocols to protect the quantum coherence even in systems coupled to a fluctuating environment \cite{Celardo2014b}. 
\item The segregation of specific solutions with a spatial patterns that can mostly match the collector, provides an emerging mechanism that enahnces the \textbf{directionality} of the transport process \cite{Scholes2011}. For example, it has been shown that FMO complex may work as a rectifier for unidirectional energy flow from the peripheral light-harvesting antenna to the reaction center complex by taking advantage of quantum coherent effects that drive an efficient trapping by the pigments facing the reaction center complex \cite{Ishizaki2009b}.
\item Energy mismatches in disordered materials lead to destructive interference of the wavefunction and subsequently to \textbf{localization} of the quantum wavefuntions \cite{Anderson1958}, which is in general detrimental to the transport efficiency. Nevertheless, if the static disorder is sufficiently strong, 
quantum coherence effects may significantly enhance transport in open systems even in the
deep classical regime (i.e. where the decoherence rate is greater than the inter-site hopping amplitude) \cite{Zhang2016}.
\item The interaction with an external environment is not always detrimental to transport. Indeed, as suggested by natural photosynthetic bio-systems, quantum coherence may be enhanced or regenerated by the interaction with the environment \cite{Mohseni2008,Rebentrost2009,Chin2010,Huelga2013}. Furthermore, the dephasing induced by the \textbf{environmental fluctuations} can greatly enhance the transport efficiency for it counteracts the coherence-driven localization. The coupling to a common decay channel, such as the collector, has been also found to promote an open-assisted energy transfer in quantum networks and light-harvesting complexes \cite{Celardo2012,Zhang2016}. 

In general to find the optimal dephasing is a difficult task. The Goldilocks principle has been proposed as a guiding principle to find the optimal balance between disorder and dephasing \cite{Lloyd2011}. The idea is the following: due to static disorder an initial excitation localized on one site will spread ballistically up to the localization length, on a typical time scale: the localization time scale. After that time, the excitation spreading will stop and the excitation will be localized. While ballistic spreading is good for transport, localization is obviously not. On the other side, dephasing induces a diffusive spreading of the excitation. The diffusive regime is better than the localized regime for transport, but it is worse than the ballistic regime. Thus the optimal dephasing should be such that the dephasing time matches the localization time, so that the excitation is free to spread up to the localization length ballistically and then dephasing can free it, allowing to continue the spreading.
\end{itemize}

\begin{figure}
\begin{center}
\includegraphics[width=1\textwidth]{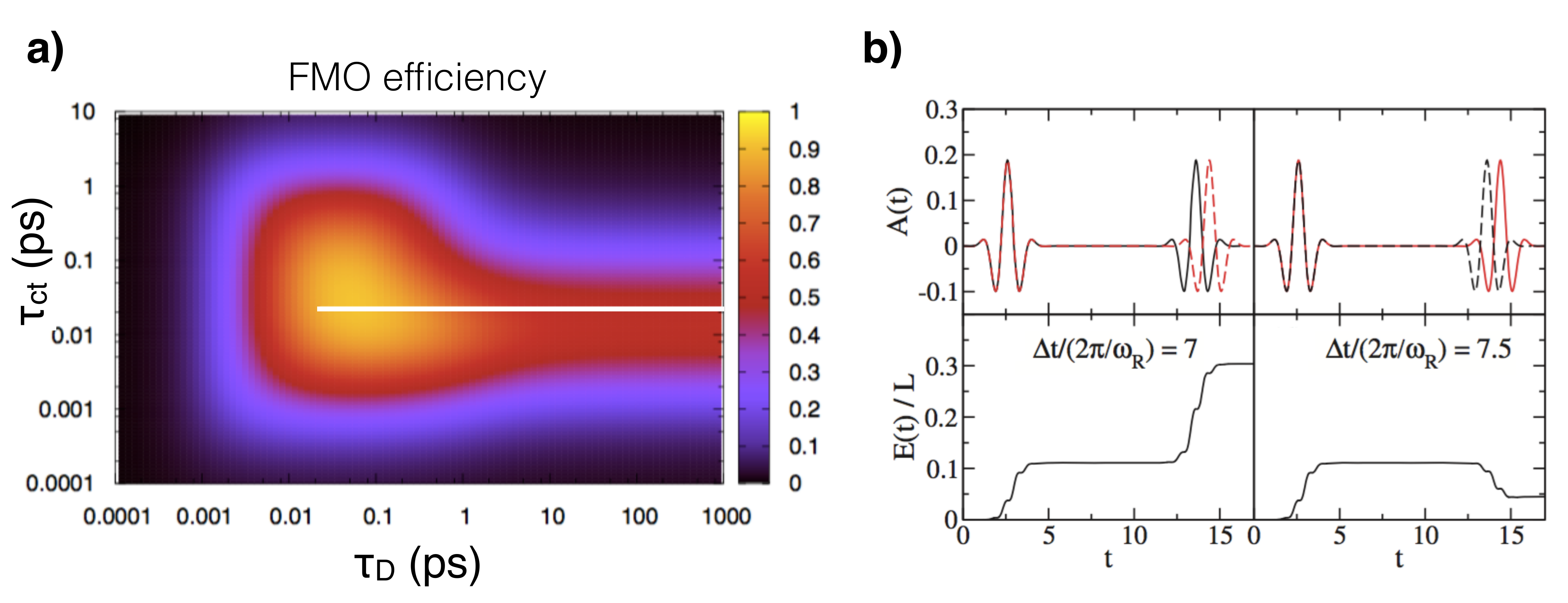} 
\caption{a) Transport efficiency at t=5 ps in the FMO system as a function of the decoherence time $\tau_D$ and the coupling to the continuum time $\tau_{ct} = \hbar/\gamma_{ct}$.
The horizonthal white line represents the superradiant transition. b) Time evolution of the vector potential (\textbf{A}(t), top panels) during the interaction of two consecutive phase-coherent pulses, whose delay can be externally controlled, with an extended Hubbard model \cite{Lu2013}. The time delay is expressed in the unit of 2$\pi$/$\omega_R$, $\omega_R$ being the Rabi frequency of the system. The bottom panel display the evolution of the total energy of the system, that can be either enhanced or reduced for different delays. Taken from Ref. \cite{Lu2013}.}
\label{fig_coherent_transport}
\end{center}
\end{figure}

\subsubsection*{\textbf{Transition-metal oxides heterostructures: taking inspiration from nature}}
The extension of the concepts discussed in the previous section to solid state systems would pave the way to the observation and exploitation of quantum-coherent phenomena in materials for technological applications. An interesting example is given by the theoretical investigation of the interaction of two consecutive coherent light pulses with a one-dimensional Mott insulator \cite{Lu2013}. The formation  of bound states of charge carriers, i.e., holon-doublon excitons, is included in the model by considering an extended Hubbard model in which, besides the on-site Coulomb repulsion $U$, nearest neighbor electron-electron interactions ($V$) are also included. The main result is that, when the decoherence time of optically active excitations is of the order of the pulse length, the many-body system can be mapped onto an effective two-level system, and its dynamics can be rationalized in terms of a simple Rabi model \cite{Lu2013}. As a consequence, the final energy after the interaction with the second pulse can be either enhanced or quenched depending on the time delay between the first and second pulses, as shown in Fig. \ref{fig_coherent_transport}b. This effect, shown in Fig. \ref{fig_coherent_transport}, is the counterpart of the quantum-coherent time evolution of a two-level system interacting with an electro-magnetic field, which is described by the optical Bloch equations \cite{Yariv}. This result suggests the possibility of coherently controlling the electronic state of a correlated material by suitable excitation with light pulses as short as $\tau_D$.

More in general, TMOs provide the ideal platform to extend the concepts of superradiance, directionality, localization and noise-assisted transport that have been introduced for addressing the transport properties of natural light-harvesting systems \cite{Huelga2013}. The starting point is that the energy of the ubiquitous charge-transfer transition usually matches most of the visible spectrum (1-3 eV) and can be thus used to directly probe and control the fundamental electronic decoherence processes on the sub-10 fs timescale \cite{Brida2010}. Furthermore, correlated materials are characterized by the unique property that the dynamics of the quasiparticles at the Fermi level and of the high-energy excitations involving the Hubbard band (the CT transition in copper oxides) are strongly intertwined \cite{Giannetti2016b}. As a consequence, the decoherence dynamics of holon-doublon excitons is expected to be strongly affected by the onset of ordered states (either local or long-ranged), such as magnetic or charge-ordered phases. Therefore, an interesting and completely unexplored mechanism that could help in tailoring the ultrafast electronic decoherence in solids is the possibility of tuning the fluctuations of the thermal bath in the vicinity of a phase transition. In this regime the relevant incoherent fluctuations (spin, phonons, charges) transform into a phase-coherent condensate, such as  magnetic, charge- and orbitally-ordered states or high-temperature superconductivity.
Besides the example of LVO, in which the orbital quantum fluctuations in the high-temperature orthorombic phase condense into an orbitally ordered states at 140 K (see section \ref{sec_transport_hetero}), another paradigmatic example of the potentiality of TMOs is given by the Nd$_{2-x}$Ce$_x$CuO$_4$ family of copper oxides. When tuning the electron doping concentration ($x$), it is possible to move away from the antiferromagnetic insulating phase and explore interesting states such as: i) bad metallic phase with long-range antiferromagnetic correlations \cite{Motoyama2007}; ii) insulating or bad metallic states with strong short-range antiferromagnetic fluctuations \cite{Motoyama2007}; iii) short-range (coherence length $\simeq$2 nm) and fluctuating charge order \cite{daSilva2015}, in which the electronic charges spontaneously tend to break the translational symmetry  and form phase-ordered patterns; iv) unconventional superconductivity at high temperatures, in which the condensation of the charge carriers into a phase coherent condensate induces the opening of a $\simeq$40 meV gap with $d$-wave symmetry. 

Wrapping up the concepts discussed in this section, we argue that TMOs provide the unique possibility of using the physics of correlated materials to tune the decoherence processes in atomically-thin heterostructures and devices. The exploitation of noise-assisted quantum-coherent phenomena such as super-transfer \cite{Lloyd2010} and superradiant transitions \cite{Celardo2012} to increase the performances and efficiency of TMO-based devices would dramatically impact the current technology. However, some fundamental steps are still required to transform these opportunities into a real technological breakthrough:
\begin{itemize}
\item \textit{Tuning the decoherence}. Coherent transport phenomena emerge in a limited region of the phase-space of the parameters $\tau_D$ and $\gamma_{ct}Q$. In order to control and tune quantum-coherent processes, it is necessary to develop protocols to span the entire phase space and continuously move from the incoherent to the coherent regime. The most promising strategy is to tune $\tau_D$ by exploiting the modification of the electronic coupling with the bosonic bath in the vicinity of a phase transition. In this perspective, TMOs provide a very interesting playground since they are always on the verge of multiple phase transitions as a consequence of the local correlations of the metal $d$ electrons. By finely tuning the temperature and doping, it is possible to turn strongly incoherent fluctuations of the charge, spin, lattice and orbital degrees of freedom into long-range ordered (and phase-coherent) states, such as magnetic phases, charge/orbital ordering and superconductivity, at temperatures ranging from 20 K to 300 K. 
\item \textit{Designing effective collectors.} One of the crucial parameters for the emergence of superradiant solutions in quantum networks is the coupling to the collectors, $\gamma_{ct}Q$. The state-of-the art technologies for the synthesis of oxide heterostructures \cite{Hwang2012} enabled the layer-by-layer control of the chemical/physical properties of nanoscaled devices. For example, the spontaneous emergence of conductive layers at the interface of insulating oxides, such as LaAlO$_3$/SrTi$_3$ and LaVO$_3$/SrTi$_3$, provide a simple scheme for the design of two-dimensional collectors that reside few atomic layers apart \cite{Huijben2006} and are characterized by high carrier mobility. Furthermore, the density of the interface conductive layers can be changed by controlling their mutual distance and their common temperature \cite{Huijben2006,Brinkman2007}, thus providing additional degrees of freedom for tuning the coupling to the collectors.
\item \textit{Tuning the hopping}. The hopping probability, $t_{\mathrm{h}}$ has a twofold role in controlling the quantum-transport properties of few atomic-layers devices. On one hand, $t_{\mathrm{h}}$ determines the fundamental timescale ($\hbar/t_{\mathrm{h}}$) of the charge-hopping process between two adjacent layers; on the other hand, it controls the mean level spacing ($\Delta$) of the energy levels of the closed quantum network with respect to the energy of the non-interacting sites. Interestingly, $t_{\mathrm{h}}$ strongly depends on the lattice structure of the material and on the orbital occupation of the TM $d$-bands \cite{Raychaudhury2007}, which can be controlled either statically, e.g. by tuning the temperature or the chemical composition, or dynamically, e.g. by resonant photoexcitation of interband transition or lattice modes \cite{Forst2011}.
\item \textit{Disorder}. The local disorder is expected to strongly impact the transport efficiency in the coherent regime, up to the point of driving the complete charge localization \cite{Anderson1958}. The disorder can be easily controlled by exploiting the intrinsic presence of defects in TMOs or by creating suitably designed disordered structures that could be controlled at the atomic scale.
\item \textit{Developing predictive models}. A key step is the development of predictive models for the description of coherent photodissociation and charge-collection processes in correlated oxides. The spreading of photo-excited carriers in Mott insulating heterostructures has been recently studied by means of DMFT solutions of the inhomogeneous Hubbard model \cite{Eckstein2013}. While DMFT fully treats the local correlations of the TM $d$-orbitals and the quantum nature of the many-body wavefunctions, only short-range spin fluctuations can be included as source of dissipative coupling to the environment. The coupling to phonons (restricted to specific dispersionless optical modes) could be included by means of the Hubbard-Holstein model. Considering the relevance of the coupling to the thermal bath and given the necessity of simulating in a simple way the spectral modifications of the noise, e.g. the transformation from an incoherent fluctuating bath to a phase-locked coherent state in the vicinity of charge- and orbitally-ordering phase transitions, the development of effective models based on appropriate master equations that can describe both coherent and incoherent hopping \cite{Ishizaki2009} is expected to boost the research in this field. The modeling of quantum-coherent processes in oxide heterostructures will guide the optimization of the experimental conditions necessary to observe and study coherent phenomena and will lead to the exploitation of these effects to improve the efficiency of transport in nanoengineered devices. 

\end{itemize}

\subsection{Multi-dimensional spectroscopies to access quantum decoherence}
\label{2Dspec}
\begin{figure}
\begin{center}
\includegraphics[width=1.2\textwidth]{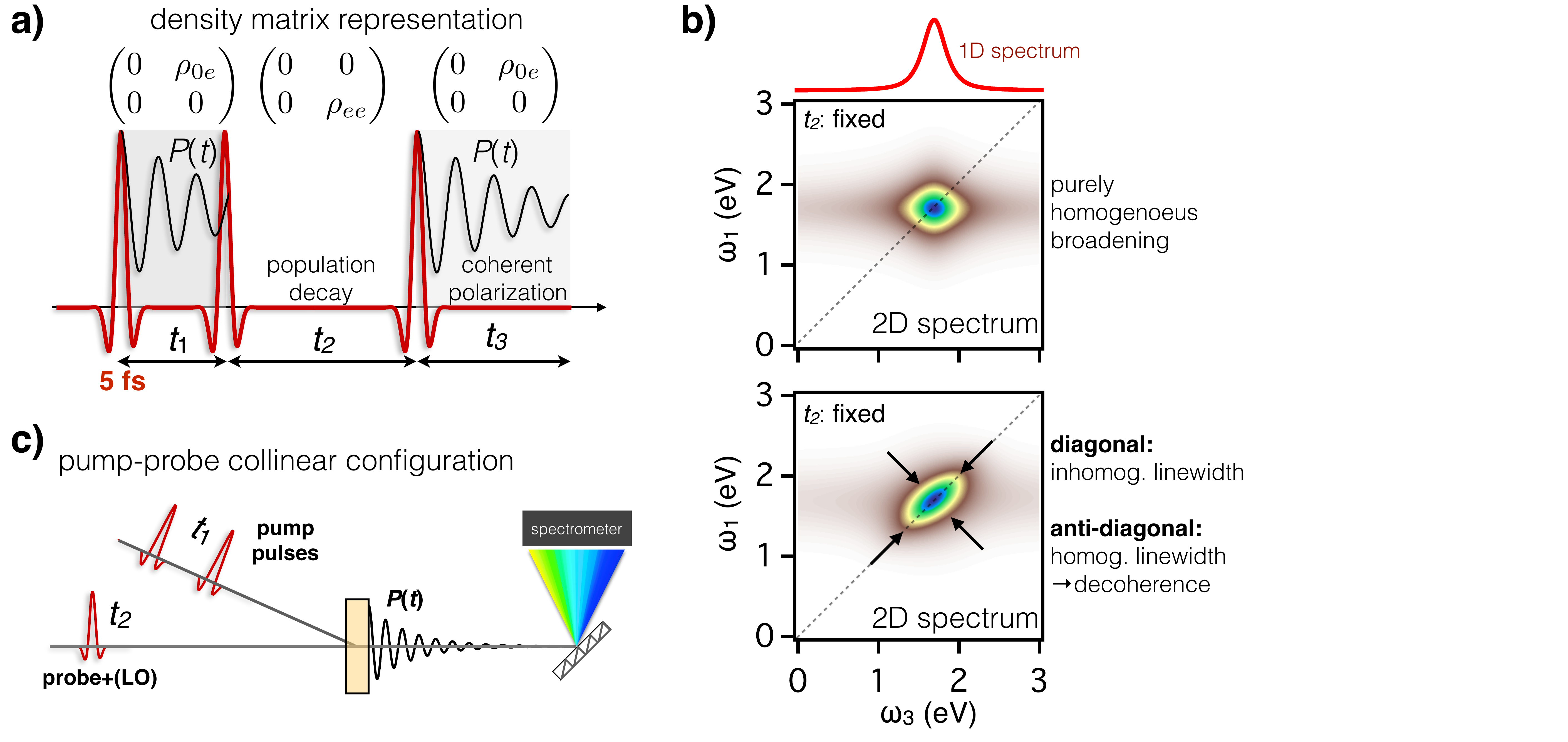} 
\caption{a) Cartoon of the 2D spectroscopy working principles. b) The panels report simulated 2D absorptive spectra for a 0.4 eV broad linewidth, representing the typical charge-transfer exciton in transition-metal oxides. In the top panel the linewidth is entirely related to homogeneous broadening. In the bottom panel, the inhomogeneous broadening is the main mechanism that determines the total linewidth. The top red curve represents the corresponding information that would be provided by a broadband pump-probe experiment. The experimental resolution (5 fs) is taken into account. c) Partially collinear pump-probe geometry for 2D spectroscopy. The probe pulse is self-heterodyned and is dispersed onto a spectrometer. Taken from Ref. \cite{Giannetti2016b}.}
\label{fig_2D}
\end{center}
\end{figure}
Up to now, any attempt to investigate the electronic dephasing in solids at high-temperatures has been hindered by the difficulties in directly measuring the fundamental dephasing time of photoexcitations. For example, the linewidths measured by conventional optical spectroscopy are strongly affected by inhomogeneous broadening, which reflects the statistical distribution of impurities. This effect is particularly severe in transition metal oxides, in which the strong correlations lead to intrinsic electronic and structural inhomogeneities. On the other hand, time-domain optical spectroscopies with ultrafast resolution are emerged as a well-established tool to investigate the dynamics of correlated oxides \cite{Giannetti2016}. Unfortunately, state-of-the art ultrafast spectroscopies simply probe the population of a specific transition and remain blind to the fundamental decoherence processes. As a consequence, the relevant mechanisms that lead to the ultrafast loss of quantum coherence of charge excitons in metal oxides and the interplay between their dephasing and the onset of long-range ordered phases is still a major unexplored field. 

In order to access the fundamental decoherence (or dephasing) processes described in the previous sections, it is mandatory to develop a technique with the following features:
\begin{itemize}
\item[-] a time resolution as short as the typical decoherence timescale (few femtoseconds) 
\item[-] capability of disentangling the real polarization dephasing from the population decay and from inhomogeneous effects
\end{itemize}
One of the most promising approaches that could fulfill these requirements is the multi-dimensional spectroscopy \cite{Hamm2011,Cundiff2013}, which was developed over the past two decades in physical chemistry and atomic physics and is the optical analogue of NMR used to study the decoherence of atomic wavefunctions and the energy transfer between vibrating molecules. 

In particular, two-dimensional (2D) spectroscopy directly accesses $\rho_{0e}$, being based on the coherent interaction of three light pulses mediated by the third-order susceptibility, $\chi^{(3)}$, of the material \cite{Hamm2011,Cundiff2013}. In the simplest picture (see Fig. \ref{fig_2D}a), the first pulse creates a macroscopic polarization, leaving the system in the coherent superposition of $\vert\psi_0\rangle$ and $\vert\psi_e\rangle$. The polarization oscillates in time with frequency $\omega_{0e}$ and decays on the dephasing timescale $\tau_D$. The second pulse, at time t$_1$, converts the macroscopic polarization into a population in the excited state, which does not oscillate but is reminiscent of the initial phase of the polarization. The third pulse, at time t$_2$ converts back to a coherent polarized state that radiates the measured signal field at time t$_3$. The two-dimensional spectrum (see Fig. \ref{fig_2D}b) is generated, for each fixed t$_2$, by scanning t$_1$ and t$_3$ and taking the Fourier transforms. 
2D spectroscopy is considered the ultimate non-linear technique \cite{Mukamel2015}, since it contains all the information that can be retrieved by 0D (conventional optics) and 1D spectroscopies, such as single- and multi-colour pump-probe and multi-pulse techniques without the probe-frequency resolution \cite{Yusupov2010,DalConte2012,Madan2014}. On the other hand, 2D spectroscopy also contains fundamental information that is absent in 0D and 1D, such as:
\begin{itemize}
\item the possibility of determining whether two resonances are coupled or uncoupled \cite{Hamm2011}
\item the capability of discriminating between homogeneous and inhomogeneous linewidths \cite{Hamm2011}
\item the possibility of reconstructing the exciton migration and the charge-separation process \cite{Abramavicius2010}
\end{itemize}
This technique already showed that quantum-phase coherence is preserved in biological systems immersed in room-temperature baths \cite{Engel2007,Collini2010,Scholes2011} and unveiled many-body excitonic effects in semiconductors at low temperature (10 K) \cite{Zhang2007}. The possibility of extending 2D spectroscopy in the X-ray and in the THz frequency domain is expected to open new routes for the coherent spectroscopy of spins, valence electrons, and and core electronic excitations.

So far, the use of 2D spectroscopy for the study of solid state systems has been mainly hindered by the limited temporal resolution of most setups ($\approx$100 fs) and by the low repetition rate (1 kHz) that forces the experiments to be performed at very high fluences and on semitransparent samples. However, the recent dramatic advances in fiber laser technology \cite{Nillon2014,Liebel2014} made available high-repetition rate sources (up to 2 MHz) with enough energy per pulse for producing and amplify supercontinuum white-light. The use of suitably designed non-collinear optical parametric amplifiers (NOPA) allows to produce ultrashort broadband pulses with tunable frequency and repetition rate and whose frequency content can support pulses as short as 6 fs \cite{Nillon2014,Liebel2014}. To show the advantages of 2D spectroscopy with respect to broadband pump-probe techniques \cite{Giannetti2016}, we simulated the absorptive \cite{Hamm2011} 2D spectra (see Fig. \ref{fig_CT}a) expected for the typical CT excitations in correlated oxides, taking into account the expected temporal resolution of the experiment (6 fs). Assuming a total linewidth of 400 meV (see Fig. \ref{fig_CT} for the actual optical conductivity of specific materials), the inhomogeneously (bottom panel) and homogeneously (top panel) broadened cases are clearly disentangled in the 2D spectra. In contrast, any 1D (or 0D) optical spectroscopy would only access the integral with respect to $\omega_1$ of the 2D spectrum and would not be able to discriminate between the two cases. 

In light of the concepts presented in this section, we foresee that the achievement of a temporal resolution of $\approx$6 fs, which is of the order of the decoherence timescale in correlated oxides, in high-repetition rate 2D experiments would unlock the study and, possibly, the control of the quantum-coherent dynamics in oxide heterostructures.
The possibility of exploiting quantum-coherent processes at ambient temperatures would stimulate the design of oxide heterostructures and nano-materials that preserve the quantum-coherence of the photo-stimulation and collect the photo-generated charges along quantum walks. 
The success in coherently converting photons into charge excitations in oxide-base devices would provide the building block to develop robust quantum systems that can be manipulated at ambient conditions, for future use in large-scale quantum-information processes and communication technologies \cite{Awschalom2007}.
These ideas and techniques could be naturally applied to a broad variety of functional oxides (e.g. manganites, vanadates, nickelates, iridates and iron oxides) that exhibit optically-driven insulator-to-metal transitions, multiferroicity, magnetic superconductivity and to low dimensional solids (e.g. graphene and metal dichalcogenides).

\section{Wave-like temperature propagation}
After a sudden quench or other types of impulsive excitation, the excess energy initially stored into the non-thermal electronic distribution is eventually shared among the fermionic and bosonic (e.g. phonons, magnetic fluctuations) degrees of freedom. On timescales longer than the time necessary for the internal thermalization, it is possible to define an effective temperature, larger than the initial one, that labels the thermodynamic distributions. However, when the typical dimensions of the system are scaled to the nanosize, the initial excitation can spread within the material faster than the internal thermalization time, thus leading to an unconventional description of the heat propagation process \cite{Joseph1989,Banfi2010}. Although the development of a microscopic model of this phenomenon is one of the challenges of nano-thermodynamics (with possible impact on the physics of granular media and liquid helium) \cite{Joseph1989}, here we will treat the problem empirically by introducing a delay between the onsets of the heat flux, $\vet{q}$ and of the temperature gradient, $\nabla T$. For sake of clarity we will briefly review the steps necessary to obtain the generalized heat propagation equation that can be applied to tackle the thermal problem at the nanoscale.    

\subsection{Generalized equation for the temperature field}
In general, the balance of the energy exchanges is rationalized through a local \emph{conservation of energy} equation:
\begin{equation}
C \frac{\partial T}{\partial t}\left(\vet{x},t\right)=-\nabla \cdot \vet{q}\left(\vet{x},t\right)+S\left(\vet{x},t\right),
\label{energy_conservation}
\end{equation}
where $C$ is the \emph{volumetric heat capacity} and $S\left(\vet{x},t\right)$ is the \emph{volumetric heat source}~\cite{Tzou} externally provided. The classical Fourier's law is obtained by assuming a direct proportionality, at time $t$, between the heat flux and the temperature gradient, i.e., $\vet{q}\left(\vet{x},t\right)=-\kappa_T\ \nabla T \left(\vet{x},t\right)$, $\kappa_T$ being the thermal conductivity. While the Fourier's principle is a good approximation for describing slow phenomena, it implicitly assumes that the heat flux and the temperature gradient are simultaneous, which implies an infinite speed of the heat propagation \cite{Tzou}. 
When considering fast timescales and small length-scales, two different scenarios can be analyzed:
\begin{itemize}
\item[i]
After the instantaneous establishment of the temperature gradient, a non-zero time, $\tau_q$, is necessary to generate the heat flux. This delay is related to the microscopic interactions that lead to a very fast \textit{local} thermalization, while the heat flux is delayed by specific dynamical constraints. A typical example is given by strongly anisotropic materials, such as iridium or copper oxides, in which the in-plane thermalization is driven very quickly by the coupling with spin fluctuations and phonons, while the heat transport perpendicular to the TMO planes is strongly quenched by the small interlayer coupling. In this case, the constitutive equation relating the heat flux and the thermal gradient is modified into the Cattaneo-Vernotte model \cite{Cattaneo,Tzou}: $\vet{q}\left(\vet{x},t+\tau_q\right)=-\kappa_T\ \nabla T \left(\vet{x},t\right)$.
\item[ii]
In the opposite limit, a heat flux is instantaneously established causing a temperature gradient at the delayed time $\tau_T$. This condition, represented by the relation $\vet{q}\left(\vet{x},t\right)=-\kappa_T\ \nabla T \left(\vet{x},t+\tau_T\right)$, is attained, for example, when a fast ballistic propagation of the charge carriers anticipates the local thermalization of the system.
\end{itemize}

By combining the two previous limits into a single relation, it is possible to introduce the Dual-Phase-Lag (DPL) model \cite{Tzou}, that describes the general constitutive relation between the heat flux and the temperature gradient: 
\begin{equation}
\vet{q}\left(\vet{x},t+\tau_q\right)=-\kappa_T\ \nabla T \left(\vet{x},t+\tau_T\right).
\label{DPLM_law}
\end{equation}

By substituting the first order term of the Taylor series in the time variable of  equation \ref{DPLM_law} into the energy conservation law at time $t$ (equation \ref{energy_conservation}) and assuming $S(\vet{x},t)$=0, it is possible to obtain the generalized partial differential equation for the temperature field:
\begin{equation}
\left( {\tau_q\over \alpha}\right)\frac{\partial^2 T(\vet{x},t)}{\partial t^2}-\nabla^2 T(\vet{x},t)+{1\over \alpha}\frac{\partial T(\vet{x},t)}{\partial t}-\tau_T\frac{\partial \nabla^2T(\vet{x},t)}{\partial t}=0,
\label{temperature_wave_equation}
\end{equation}
where $\alpha=\kappa_T/C$ is the \emph{thermal diffusivity}.
Interestingly, equation \ref{temperature_wave_equation} recalls a wave equation in which $\sqrt{\alpha/\tau_q}$ is the speed of propagation and $\sqrt{\alpha\tau_q}$ represents the diffusion length. The damping of the temperature wave is provided by a diffusive term and a third-derivative one. Eq. \ref{temperature_wave_equation} reduces to the conventional diffusion equation in the limit $\tau_q$,$\tau_T\rightarrow$0, which implies and infinite speed of the heat propagation.
From Eq. \ref{temperature_wave_equation} it is possible to extract the complex dispersion relation of the system, linking the wavevector $k$=$|\vet{k}|$ to the frequencies ($\omega$) of the propagating temperature waves: 
\begin{equation}
k^2\left(1+i\omega \tau_T\right)=\left( {\tau_q\over \alpha}\right)\omega^2\left(1-\frac{i}{\omega \tau_q} \right).
\label{complex_dispersion}
\end{equation}

\subsection{Wave-like solutions}
\label{wave_like_solutions}

Oscillating solutions of Eq. \ref{temperature_wave_equation} can be sought under the assumption that $k$ and $\omega$ entering in Eq. \ref{complex_dispersion} are complex quantities. Having in mind the experimental configuration of ultrafast experiments, in which the laser impinges on one of the sides of the sample and acts as an initial volumetric heat source $S(\vet{x},\mathrm{t}$=0), we are interested in studying how the given initial temperature profile evolves in time. As a consequence, we assume a real wavevector, while the frequency is given by the complex quantity $\tilde{\omega}=\omega_1+i\omega_2$, where $|\omega_1|$=2$\pi$/$t_{osc}$ is the inverse of the oscillation period and $\omega_2$=1/$t_{damp}$ is the inverse of the damping time. The onset of a wave-like regime is governed by the $Q$-factor $Q$=$|\omega_1|$/$\omega_2$, which discriminates the overdamped ($Q\ll$1) from the underdamped regimes ($Q\gg$1). For simplicity, we will consider the one-dimensional problem in which the spatial coordinate, $z$, is perpendicular to the sample surface.  

By substituting the complex frequency $\tilde{\omega}$ into the dispersion \ref{complex_dispersion}, it is possible to obtain the analytic expressions of $\omega_1$ and $\omega_2$ as a function of $k$. Without entering into the details of the calculations, that will be presented elsewhere \cite{Gandolfi2016}, it is possible to show that the nature of the admitted solutions strongly depends on the ratio $\tau_T$/$\tau_q$. While non-oscillatory solutions, characterized by $\omega_1 =$0, are found for $\tau_T \geq \tau_q$, i.e. when the heat flux precedes the establishment of a thermal gradient, a wave-like behaviour of the temperature propagation may emerge in the case $\tau_T<\tau_q$, i.e., when the temperature gradient is established before the onset of the heat flux. We can define a lower ($k_{low}$) and upper ($k_{up}$) bound for the wavevectors that can sustain oscillatory solutions:
\begin{equation}
\label{k_boundaries}
k_{low(up)}=\sqrt{\frac{2}{\alpha \tau_T}\left(\frac{ \tau_q}{ \tau_T}\right)\left(1-\frac{1}{2}\frac{\tau_T}{\tau_q}-(+)\sqrt{1-\frac{\tau_T}{\tau_q}}\right)}.
\end{equation} 
In the range $k_{low}<k<k_{up}$, the allowed complex frequencies are:
\begin{equation}
\label{omega_1}
\fl \omega_1 =\mp\sqrt{-\left[\frac{\alpha^2}{4}\left(\frac{\tau_T}{\tau_q}\right)^2 k^4+\frac{\alpha}{\tau_q}\left[{1\over2}\left(\frac{\tau_T}{\tau_q}\right) -1\right]k^2+\frac{1}{4\tau_q^2}\right]} ;\\  \omega_2=\frac{1}{2\tau_q}+\frac{\alpha}{2}\left[\frac{\tau_T}{\tau_q}\right] k^2\\
\end{equation}
while, for $k\leq k_{low}$ and $k\geq k_{up}$ no oscillatory solutions are admitted, i.e. $\omega_1 = 0$.\\

The oscillating behaviour of the temperature field, is better appreciated by the $Q$-factor:
\begin{equation}
Q(k)=\left\{\begin{array}{ll}
\displaystyle{\sqrt{\frac{4\alpha \tau_q k^2}{\left(1+\alpha\tau_T k^2 \right)^2}-1\ }}& \ \ \ \ for\  k_{low}<k<k_{up}\\
\\
0& \ \ \ \  for\ k\leq k_{low}\ \vee\ k\geq k_{up}.\\
\end{array} \right.
\label{Q_factor_TauqmagTauT}
\end{equation}
The most favorable condition to observe underdamped oscillations corresponds to the maximum $Q$-factor:
\begin{equation}
Q_{max}=\sqrt{{\tau_q\over \tau_T}-1},
\label{Qmax}
\end{equation}
that is obtained at the wavevector $k_{Q_{max}}=1/\sqrt{\alpha\tau_T}$.
Equation \ref{Qmax} suggests that the condition $\tau_q\gg\tau_T$ is the prerequisite to observe wave-like effects of the temperature propagation in real materials.

\begin{figure}
\begin{center}
\includegraphics[width=1\textwidth]{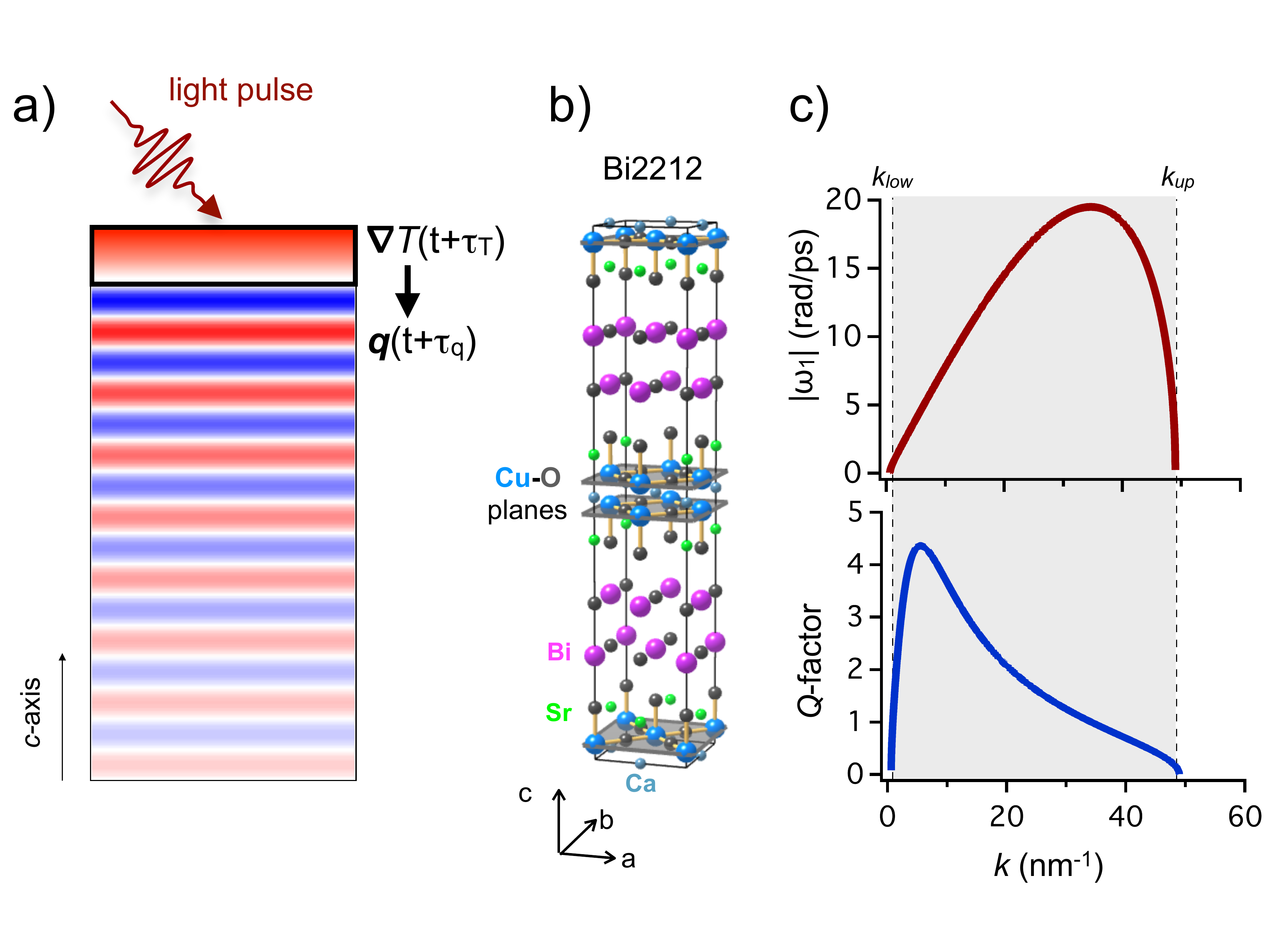} 
\caption{a) Cartoon of the generation of wave temperatures on ultrafast timescales. A temperature gradient is quickly established ($\tau_T\simeq$50 fs), while the heat flux follows at time $\tau_q>$1 ps. b) Unit cell of Bi$_2$Sr$_2$CaCu$_2$O$_8$. The Cu-O planes are highlighted by the gray surfaces. c) Real part of the eigenfrequencies ($|\omega_1|$) and $Q$-factor, as calculated by Eqs. \ref{omega_1} and \ref{Q_factor_TauqmagTauT}, with the values $\alpha$=6.5$\cdot$10$^{-7}$ m$^2$/s, $\tau_q$=1 ps and $\tau_T$=50 fs, corresponding to the case of electronic conduction at $T$=20 K in Bi2212 along the $c$-axis. $k_{low}$ and $k_{up}$ are indicated by the dashed lines.}
\label{fig_waveT}
\end{center}
\end{figure}

\subsection{Emergent phenomena at the nanoscale}
As shown in Sec. \ref{wave_like_solutions}, emergent wave-like regimes of the temperature field may potentially be found in systems in which the condition $\tau_q\gg\tau_T$ holds, i.e., a thermal gradient is very quickly established, while the heat flux follows at a later time $\tau_q$.\\ Correlated TMOs provide a very interesting playground to investigate this physics.
Their layered structure is characterized by strongly anisotropic conduction properties that can lead to a very fast in-plane thermalization and to a very slow out-of-plane heat propagation, thus favoring the emergence of the wave-like regime, as schematically shown in Fig. \ref{fig_waveT}a.\\

Here, we will focus on copper oxides and, in particular, on the Bi$_2$Sr$_2$CaCu$_2$O$_8$ (Bi2212) family that exhibits the most anisotropic transport properties. In general, the conduction properties of copper oxides are dominated by the charge carriers that occupy the Cu-3$d$ orbitals within the Cu-O planes ($ab$-planes, see Fig. \ref{fig_waveT}b). The strong tendency to delocalization, driven by the very large in-plane hopping integral ($t_{\parallel}\sim$0.3 eV) is contrasted by the strong effective interactions. At moderate doping concentration ($\sim$15\%) the electronic interactions give rise a strong coupling with the bosonic fluctuations of both magnetic and phononic origin \cite{DalConte2012}. The fastest scattering process is likely to be related to the coupling with the short-ranged antiferromagnetic fluctuations that persist when the system is doped \cite{LeTacon2011}. It has been recently shown \cite{DalConte2015} that the excess kinetic energy provided by the pump pulse is completely released to the magnetic background within a timescale of $\sim$50 fs. \\
It is thus reasonable to assume $\tau_T$=50 fs as the time necessary for the local thermalization between charge carriers and spin fluctuations and for the onset of the temperature gradient after an anisotropic pump excitation.\\ 
The electron-phonon coupling eventually drives the complete thermalization between the electronic, magnetic and phononic degrees of freedom on a timescale of $\sim$1 ps.
We underline that, even in bulk materials, the pump excitation is intrinsically anisotropic, being the energy absorption confined within a penetration length of $\sim$20-200 nm depending on the material. As we will discuss later, "artificial" gradients can be induced by nanostructuring the material under study.

\begin{table}
\caption{\label{table1} Electronic heat capacity ($C$), thermal conductivity ($k_T$) and thermal diffusivity ($\alpha$) of copper and iridium oxides. $C$ and $k_T$ of nearly optimally doped ($T_c$=85 K) Bi2212 crystals are taken from Refs. \cite{Crommie1991}, \cite{Junod1994} and \cite{Loram2000}. These values are representative of the heat capacity and thermal conductivity of other families of copper oxides, such as YBa$_2$Cu$_3$O$_7$ and La$_{2-x}$Sr$_x$CuO$_4$. $C$ and $k_T$ of Sr$_2$IrO$_4$ crystals are taken from Ref. \cite{Kini2006}. These values are representative of the heat capacity and thermal conductivity of other families of iridates, such as Na$_2$IrO$_3$.}
\vspace{0.5 cm}
\footnotesize
\begin{center}
\begin{tabular}{@{}lllllllll}
\ns
\textbf{Bi$_2$Sr$_2$CaCu$_2$O$_8$}\\
\textbf{(Bi2212)}\\
\br
&&&&&&&&diffusion\\&$T$&$C$&$k_T$&$\alpha$&$\tau_T$&$\tau_q$&velocity&length
\\&K&J/m$^3$K&W/mK&m$^2$/s&ps&ps&nm/ps&nm\\
\mr

electrons&300&3.3$\cdot$10$^4$&3.0$\cdot$10$^{-4}$&1.0$\cdot$10$^{-8}$&0.05&$>$1&$<$0.10&$>$0.10\\
&\ 50&4.0$\cdot$10$^3$&2.5$\cdot$10$^{-4}$&6.5$\cdot$10$^{-8}$&0.05&$>$1&$<$0.25&$>$0.25\\
&\ 20&4.0$\cdot$10$^2$&2.5$\cdot$10$^{-4}$&6.5$\cdot$10$^{-7}$&0.05&$>$1&$<$0.80&$>$0.80\\

\mr
\end{tabular}\\\vspace{2 cm}


\begin{tabular}{@{}lllllllll}
\ns
\textbf{Sr$_2$IrO$_4$}\\ \textbf{(SIO)}\hspace{0.3cm} \\
\br
&&&&&&&&diffusion\\&$T$&$C$&$k_T$&$\alpha$&$\tau_T$&$\tau_q$&velocity&length
\\&K&J/m$^3$K&W/mK&m$^2$/s&ps&ps&nm/ps&nm\\
\mr
electrons/\ \ \ \ \ \ \ \ &300&1.5$\cdot$10$^3$&1.5$\cdot$10$^{-7}$&1$\cdot$10$^{-10}$&0.1&$\approx$1000&$\approx$3$\cdot$10$^{-4}$&$\approx$0.3\\
magnetic&\ 50&2.5$\cdot$10$^2$&2.5$\cdot$10$^{-8}$&1$\cdot$10$^{-10}$&0.1&$\approx$1000&$\approx$3$\cdot$10$^{-4}$&$\approx$0.3\\
fluctuations&\ 20&1.0$\cdot$10$^2$&1.0$\cdot$10$^{-8}$&1$\cdot$10$^{-10}$&0.1&$\approx$1000&$\approx$3$\cdot$10$^{-4}$&$\approx$0.3\\
\br
\end{tabular}\\
\end{center}
\end{table}

Notably, the transport properties of copper oxides are  strongly anisotropic, being the interlayer hopping limited to $t_{\perp}\approx$15 meV \cite{Feng2000}. As a consequence, the strong in-plane scattering makes the $c$-axis conductivity almost completely incoherent.\\
In Bi2212, the $c$-axis optical conductivity can be as small as 0.01 $\Omega^{-1}$cm$^{-1}$, in contrast to an in-plane conductivity of the order of 5000 $\Omega^{-1}$cm$^{-1}$ \cite{Basov2005}. In a similar way, the \textit{electronic} thermal conductivity is reduced from $\sim$1 W/mK in the $ab$-plane to $\sim$3$\cdot$10$^{-4}$ W/mK when considering the interlayer thermal transport. The natural consequence of this strong anisotropy is that any possible heat exchange perpendicular to the Cu-O planes is strongly delayed as compared to the in-plane thermalization that is effective on the sub-100 fs timescale.\\
This observation naturally suggests a possible building block for designing materials in which the wave-like propagation of the temperature field can be observed and exploited. Given the difficulty of estimating $\tau_q$ from both first-principles and experiments, we will consider the value of $\sim$1 ps as a lower bound. This value, which represents a conservative estimate, corresponds to the reasonable assumption $\tau_q$/$\tau_T$=$t_{\parallel}$/$t_{\perp}$.\\
Considering the electronic problem related to the heat propagation along the $c$-axis: in Bi2212, we obtain $\alpha\sim$10$^{-8}$-6.5$\cdot$10$^{-7}$ m$^2$/s in the 300-20 K temperature range, as reported in table \ref{table1}, while $\tau_q$/$\tau_T>$20. From Eq. \ref{k_boundaries} and assuming $\tau_q$=1 ps, it is possible to calculate the upper ($k_{up}\sim$400-50 nm$^{-1}$ at $T$=300-20 K) and lower ($k_{low}\sim$5-0.63 nm$^{-1}$ at $T$=300-20 K) bounds of the wavevectors compatible with temperature oscillations. While 2$\pi$/$k_{up}$ corresponds to a fraction of the interatomic distance and cannot be experimentally achieved, $\lambda_T$=2$\pi$/$k_{low}$=1-10 nm provides the maximum wavelength that can sustain an oscillatory behavior. These values are compatible with the geometric constraints that can be applied by nanostructuring copper oxides and other transition-metal oxides. In Fig. \ref{fig_waveT}c we report the dispersion relation of the oscillating solution at $T$=20 K, as calculated by Eq. \ref{omega_1}. At the wavevector $k\sim$5 nm$^{-1}$, corresponding to $\lambda_T\sim1.2$ nm, the oscillation frequency is $\omega_1\sim$4 rad/ps and the $Q$-factor is compatible with the observation of oscillations with period of $\approx$1.6 ps. 

The physics discussed in the current section suggests that it is possible to design realistic nanostructured systems in which, on the picosecond timescale, the temperature propagation is described by a wave equation instead of the typical diffusion law.\\
Obviously, this regime can be achieved and observed only on fast timescales \cite{Banfi2010}, i.e. before the damping washes out the coherent effects and restores the classical behavior. The possibility of controlling layer-by-layer the growth of TMOs \cite{Hwang2012} opens intriguing perspectives: by creating heterostructures with typical dimensions of few nanometers, it is indeed possible to exploit the novel wave-like regime of the temperature field. In these systems, the local temperature increase could be manipulated at the nanoscale by exploiting interference effects, thus opening an entirely new field in the nano-thermodynamics. Furthermore, the possibility of creating superlattices with periodicity of the order of few atomic cells could lead to the design of artificial bandstructures for the temperature waves, thus opening to the development of waveguides, cavities and frequency filters for the temperature field.

Far from pretending of being exhaustive, the above discussion aims at demonstrating that, for realistic materials, it is possible to investigate an emergent wave-like regime for the temperature propagation on ultrafast timescales. The numbers reported in Table \ref{table1} should be considered as just an example of the physics that could be investigated at the nanoscale. The concepts here proposed can be easily extended to many different cases. Although in the previous discussion we focused on the electronic temperature the specific copper oxide Bi$_2$Sr$_2$CaCu$_2$O$_8$, similar effects are expected in copper oxides in general or other TMOs when considering the lattice temperature.

\begin{table}
\caption{\label{table1} Lattice heat capacity ($C$), thermal conductivity ($k_T$) and thermal diffusivity ($\alpha$) of copper oxides. $C$ and $k_T$ of nearly optimally doped ($T_c$=85 K) Bi2212 crystals are taken from Refs. \cite{Crommie1991}, \cite{Junod1994} and \cite{Loram2000}. These values are representative of the heat capacity and thermal conductivity of other families of copper oxides, such as YBa$_2$Cu$_3$O$_7$ and La$_{2-x}$Sr$_x$CuO$_4$.}
\label{Phonon_data}
\vspace{0.5 cm}
\footnotesize

\begin{center}
\begin{tabular}{@{}lllllllll}
\ns
\textbf{Bi$_2$Sr$_2$CaCu$_2$O$_8$}\\
\textbf{(Bi2212)}\\
\br
&&&&&&&&diffusion\\&$T$&$C$&$k_T$&$\alpha$&$\tau_T$&$\tau_q$&velocity&length
\\&K&J/m$^3$K&W/mK&m$^2$/s&ps&ps&nm/ps&nm\\
\mr

phonons&300&2.4$\cdot$10$^6$&0.9&3.8$\cdot$10$^{-7}$&1&$>$10&$<$0.2&$>$2\\
&\ 50&5.0$\cdot$10$^5$&0.5&1.0$\cdot$10$^{-6}$&1&$>$10&$<$0.3&$>$3\\\
&\ 20&8.0$\cdot$10$^4$&0.5&6.3$\cdot$10$^{-6}$&1&$>$10&$<$0.8&$>$8\\
\br

\end{tabular}\\
%
%
%
%
%
%
\end{center}
\end{table}

For example, in Bi2212 the largest value of $\alpha$ for the phonon case (6.3$\cdot$10$^{-6}$ m$^2$/s at 20 K) leads to the possibility of observing wave-like regimes at smaller wavevectors, corresponding to length scales of the order of tens of nanometers. For sake of completeness we report the relevant data for the phonon case in table \ref{Phonon_data}.\\
Similar concepts can also be extended to the spin temperature in systems characterized by short-ranged magnetic correlations. As an example, let us consider iridium oxides, which are characterized by a strong anisotropy of the magnetic dynamics. As discussed in Section \ref{iridium_oxides}, the ratio between the in-plane and out-of-plane magnetic exchange can be as high as $J_{\parallel}$/$J_{\perp}\sim$6$\cdot$10$^4$. Under non-equilibrium conditions, the strong anisotropy leads to a very fast in-plane thermalization of the magnetic fuctuations ($\tau_{T} \sim$ 100 fs) while the heat flux perpendicular to the Ir planes is strongly delayed by the small interlayer magnetic coupling. As a conservative estimation, we assume $\tau_q$=10$^4\tau_T$ and we use the electronic heat capacity and thermal conductivity of Sr$_2$IrO$_4$ (SIO), reported in table \ref{table1}, as the upper bound of the magnetic contribution. The large $\tau_q$/$\tau_T$ turns out to be favorable to observe undulatory effects of the effective temperature of magnetic fluctuations in iridium oxides. At wavevectors compatible with the dimensions of the SIO unit cell along the $c$-axis ($\simeq$25 \AA), the oscillation frequency is about $\omega_1\simeq$5$\cdot 10^{-4}$ rad/ps, with $Q>$1. Many strategies could be devised in order to tune the wavelength and the period of the coherent oscillation of the magnetic temperature. For example, larger values of the $\alpha$ parameter could be obtained by increasing the thermal conductivity via chemical doping of the Mott insulating phase. Furthermore, the use of strongly frustrated magnetic systems in proximity of a 3D magnetic transition could greatly enhance the magnetic heat capacity. A paradigmatic system is Na$_2$IrO$_3$ that exhibits a zig-zag phase transition at $T_N$=15 K. In the 70-15 K temperature range, the heat capacity is dominated by the contribution of the short-range 2D magnetic fluctuations, giving rise to $C_{mag}\simeq$4$\cdot$10$^4$ J/m$^3$K. The dramatic changes of the magnetic heat capacity in the vicinity of $T_N$ could be exploited to manipulate the wave-like nature of the magnetic temperature propagation in heterostructures or nanoscaled devices.

\section*{Acknowledgments}
F.B acknowledge financial support from the MIUR-Futuro in ricerca 2013 Grant in the frame of the ULTRANANO Project (project number: RBFR13NEA4). 
F.B, G.F and C.G acknowledge support from Universit\`a Cattolica del Sacro Cuore through D1, D.2.2 and D.3.1 grants. F.B and G.F acknowledge financial support from Fondazione E.U.L.O.

\section*{References}
\bibliography{biblio}

\end{document}